%% file: Main_Final.tex
\documentclass[lettersize,journal]{IEEEtran}
\usepackage{amsmath,amsfonts}
\usepackage{algorithmic}
\usepackage{algorithm}
\usepackage{array}
\usepackage{textcomp}
\usepackage{url}
\usepackage{verbatim}
\usepackage{graphicx}
\usepackage{cite}
\usepackage{multirow}
\usepackage{multicol}
\usepackage{comment}
\usepackage{colortbl}
\usepackage{booktabs, multirow, array, makecell, caption}
\usepackage{adjustbox}
\usepackage{relsize}
\usepackage{scalefnt}
\usepackage{amsmath,amssymb,amsfonts}

\usepackage{textcomp}
\usepackage{xcolor}

\usepackage{subcaption}
\usepackage{float}
\usepackage{multirow}
\usepackage{eucal}
\usepackage{comment}
\usepackage{mathtools}

\usepackage{scalerel}[2016-12-29]
\def\stretchint#1{\vcenter{\hbox{\stretchto[440]{\displaystyle\int}{#1}}}}

\usepackage{multirow}
\usepackage{tabularray}

\newcommand{\VDis}{-0.2 cm}

\newcounter{example} 
\newenvironment{example}[1][]{%
	\stepcounter{example}%
	\par\vspace{5pt}\noindent
	\textbf{Example~\theexample}%
	\hspace{0.2 cm}\noindent\rmfamily}%

\newcounter{assumption} 
\newcounter{definition} 
\newcounter{problem} 
\newenvironment{problem}[1][]{%
	\refstepcounter{problem}%
	\par\vspace{5pt}\noindent
	\textbf{Problem~\theproblem}%
	\hspace{0.2 cm}\noindent\rmfamily}%

\newcounter{lemma} 
\newcounter{theorem} 
\newcounter{remark} 
\newenvironment{remark}[1][]{%
	\refstepcounter{remark}%
	\par\vspace{5pt}\noindent
	\textbf{Remark~\theremark:}%
	\hspace{0.2 cm}\noindent\rmfamily}%

\newcounter{prooff} 
\begin{document}

	\title{\scalefont{1.4} \linespread{0.8}\selectfont Optimal-coupling-observer AV motion control securing comfort in the presence of cyber attacks
	}
	
	\author{Farzam Tajdari, Georgios Papaioannou, and Riender Happee
		\thanks{This work was partially supported by HiDrive project under Grant
			91561. (Corresponding author:
			Farzam Tajdari, e-mail: f.tajdari@tudelft.nl).}
		
		\thanks{Farzam Tadjari, Georgios Papaioannou, and Riender Happee are with the Faculty of Mechanical Engineering, Cognitive Robotics group (CoR), Delft
			University of Technology, 2628CD Delft, The Netherlands.}
		
		\thanks{Farzam Tajdari is also with the
			Department of Mechanical Engineering, Dynamics and Control (D\&C) group,
			Eindhoven University of Technology, 5612AZ Eindhoven, The Netherlands.}
	}
	
	
	
	\maketitle
	
	\begin{abstract}
		The security of Automated Vehicles (AVs) is an important emerging area of research in traffic safety. Methods have been published and evaluated in experimental vehicles to secure safe AV control in the presence of attacks, but human motion comfort is rarely investigated in such studies. 
		
		In this paper, we present an innovative optimal-coupling-observer-based framework that rejects the impact of bounded sensor attacks in a network of connected and automated vehicles from safety and comfort point of view.
		We demonstrate its performance in car following with cooperative adaptive cruise control for platoons with redundant distance and velocity sensors.
		The error dynamics are formulated as a Linear Time Variant (LTV) system, resulting in complex stability conditions that are investigated using a Linear Matrix Inequality (LMI) approach guaranteeing global asymptotic stability.
		
		We prove the capability of the framework to secure occupants' safety and comfort in the presence of bounded attacks. In the onset of attack, the framework rapidly detects attacked sensors and switches to the most reliable observer eliminating attacked sensors, even with modest attack magnitudes. Without our proposed method, severe (but bounded) attacks result in collisions and major discomfort. With our method, attacks had negligible effects on motion comfort evaluated using ISO-2631 Ride Comfort and Motion Sickness indexes. The results pave the path to bring comfort to the forefront of AVs security. 
		
	\end{abstract}


\section{Introduction}
\vspace{-0.3 cm}
\subsection{Background}
\IEEEPARstart{I}{n} the realm of contemporary automotive technology, the advent of Connected and Automated Vehicles (CAVs) has become increasingly pronounced~\cite{lu2014connected, guerrero2015integration, khalil2022connected, tajdari2020feedback, tajdari2021adaptive, tajdari2019integrated}.
The distinguishing feature of these CAVs lies in their capacity to accurately sense position and velocity of other vehicles, and exchange information seamlessly, facilitating the formation of platoons~\cite{li2017dynamical, zheng2017platooning, zhou2022robust, li2023adaptive, tajdari2023online}. 
This collaborative approach enhances their efficiency in traversing distances, potentially elevating road capacity, smoothening traffic flow, curbing fuel consumption, and reducing collision rates.
However, there are still great challenges to overcome regarding the two main components of the CAVs before they are part of our daily life. 
On one hand, methods to design CAV perception and motion control struggle with securing and proving safety. On the other hand, these methods also have to ensure motion comfort, through "smooth" driving styles preventing motion sickness \cite{diels2016self,papaioannou2025occupants}. 
Securing safety refers to guaranteeing stability of the vehicles facing external factors and avoiding collisions with other vehicles, while motion comfort refers to the experience of occupants and is directly affected by acceleration and deceleration of the vehicle~\cite{aledhari2023motion} and motion relative to other road users\cite{he2024new, he2022modelling}.


CAV platoons harness information exchanged through Vehicle-to-Vehicle (V2V) communication and onboard sensors to establish informed decisions and strategically adjust the collective behavior of vehicles.
This approach ensures the stable lateral and longitudinal control of the platoon during operation. Lateral formation control~\cite{lian2022fuzzy} pertains to the lateral movement of the vehicles within the platoon, guided by the topological structure, potentially altering the platoon's overall topology.
On the other hand, longitudinal formation control~\cite{bian2021fuel, li2022adaptive} focuses on maintaining a safe following distance and velocity between vehicles, swiftly achieving internal and string stability within the platoon.
Due to the inherent instability of V2V communication networks, the longitudinal control of CAV platoons encounters numerous challenges from securing safety point of view, since potential cyber attacks can hamper vehicle stability and occupant's perceived safety and motion comfort \cite{petit2014potential, deng2021deep, adhikari2021security, nikitas2022deceitful, nordhoff2023driver, nordhoff2020using}. 
Effective cyber-attacks encompass Denial of Service (DoS)~\cite{li2023bumpless, zhao2021resilient}, False Data Injection (FDI)~\cite{biroon2021false}, replay attacks~\cite{xie2022distributed}, and hybrid cyber-attacks~\cite{lian2022fuzzy, ju2020deception}. Among these, FDI attacks, investigated in this paper, are of particular concern to researchers in the context of vehicle platoon systems~\cite{biroon2021false, yang2022resource, zhou2022attack, chong2015observability}. FDI attacks represent a prevalent and harmful form of network intrusion, disrupting the decision-making processes of communicated vehicle controllers within the platoon by intercepting and injecting misleading information into the wireless communication channel, often circumventing existing firewalls~\cite{fawzi2014secure, shoukry2014event,hassani2009multiple}. 

\vspace{-0.3 cm}
\subsection{Motivation}
In the field of AVs, motion discomfort and in particular Motion Sickness (MS) are considered, together with cybersecurity, as the main inhibitors of AVs adoption and acceptance. In vehicle design, road induced vertical vibrations, also referred to as "ride comfort" have received ample attention leading to advanced suspension systems.
However, with the advent of AVs, the interest has shifted towards horizontal accelerations, and more specifically longitudinal acceleration which occurs during accelerating and braking. 
The reason is that the longitudinal movement of automated vehicles can be influenced by motion planning, allowing the engineers to design controllers that could enhance occupants’ comfort and mitigate motion sickness, as highlighted by Elbanhawi et al.~\cite{elbanhawi2015passenger}. 
Thus, longitudinal movement in AVs and platooning vehicles draws substantial attention to potential challenges in securing safety and securing comfort. 

Based on the literature, there are multiple research works focusing on platooning or motion planning of AVs focusing on vehicle stability, or occupants' safety and motion comfort \cite{htike2021fundamentals,jain2023optimal, tajdari2019fuzzy, tajdari2021simultaneous, tajdari2020intelligent, tajdari2022flow}, while others propose methods to sort motion planner alternatives based on multiple criteria \cite{papaioannou2022multi}. 
However, to the authors' knowledge, there is no literature exploring AV control under security attacks with consideration of occupant's motion comfort. 
In this direction, this paper will explore the nuanced challenges associated with longitudinal formation control in CAV platoons under security attacks by developing a framework that is capable of enhancing occupants' motion comfort.


Due to the two main components of CAVs, only considering vehicles' safety requirements, is not sufficient \cite{limbasiya2022systematic} unless occupants' motion comfort is secured.
The traditional approaches that utilized Adaptive Cruise Control (ACC) encounter amplified shock-wave effects due to delay in sensor and actuator systems \cite{rajamani2002semi} and therefore, might be likely to cause discomfort. 
Similarly, any type of attack e.g., replay/delay attack~\cite{linkov2019human, wang2020modeling} that results in acceleration and deceleration can cause discomfort \cite{shet2019performance}.
In the absence of attack, the safety, namely collision avoidance~\cite{kuang2024research}, and comfort, namely jerk control~\cite{liu2023efficient} issues in particular in longer platoons have been resolved by communication within platoons through so-called "Cooperative Adaptive Cruise Control (CACC)"~\cite{oncu2014cooperative}. However, the impact of security attacks on occupants' motion comfort and the ability of the various cybersecurity methods to mitigate it have not yet been studied according to the authors' knowledge. 
In the occurrence of significant attacks, vehicle safety is the primary concern, namely from car crashes, but the occupants' comfort is also at risk \cite{wolf2004security, linkov2019human}. Meanwhile, even if a security method successfully reduces the impacts of the attacks, occupants' discomfort and motion sickness may be still significantly affected.

Fundamentally, there are two main categories based on the performance of existing cybersecurity methods for platooning vehicles~\cite{zhou2023robust}. 
In the first category, methods target to minimize the impact of attacks to ensure the boundedness of the closed-loop error system~\cite{chong2015observability, jin2019adaptive, yadegar2019output, jahanshahi2018attack} rather than completely eliminating the impact.
Although safety is secured, there is no guarantee that the remaining minimised attack also ensures occupants' motion comfort.
In the second category, existing approaches~\cite{li2023secure, petrillo2017collaborative, zheng2017platooning, kremer2020state, merco2018replay} identify the most reliable sensors and only use these for the CAV control, resulting in a switching system. 
However, the switching between different sensor sets usually excites the dynamic system states, which can also negatively affect the occupants' comfort.

\vspace{-0.3 cm}
\subsection{Contributions}
Accordingly, there is a lack of methods that secure robustness and safety and can guarantee the stability of the closed-loop system in the presence of bounded FDI attacks, meanwhile rejecting the impact of the attacked sensors to secure the occupants' comfort. 
Therefore, we aim to design an optimal-coupling-observer-based framework (synchronizing the observers with the most reliable observer), capable of robustly rejecting the impact of multiple sensor attacks. 
Our primary focus in this paper centers on securing longitudinal string stability and comfort for a connected homogeneous vehicles platoon in the face of bounded sensor attacks. 

The main contributions of our work are:

\begin{itemize}
    \item Presenting a novel nonlinear observer method detecting and rejecting attacked sensors, enhancing the observability of the dynamic system.
    \item Analytically deriving the observer parameters using a Linear Matrix Inequality (LMI) approach to achieve global asymptotic stability of the overall estimation error guaranteeing marginal excitation from attacks.
    \item Assessing the framework's ability to mitigate occupants' motion discomfort using ISO-2631 Ride Comfort and Motion sickness indexes, while ensuring safety in car following and platooning for different types of bounded sensor FDI attacks, such as white noise, repeatedly on-off switching white noise, stepwise attacks, and repeatedly on-off switching stepwise attacks.
    \item Benchmarking the framework against three State-Of-The-Art (SOTA) cybersecurity methods, our method detects and rejects attacks more quickly and more effectively minimises effects of attacks by excluding the attacked sensors resulting in superior safety and comfort.
    \color{black}
\end{itemize}

We present a generic optimal-coupling-observer framework with guaranteed global stability (Section~\ref{sec:methodology}), to secure any linear time-invariant system from sensor attacks. The system can include any finite number of components, e.g., any number of vehicles in platooning. The method includes only constant parameters designed off-line using an LMI approach, allowing real-time implementations. In Section~\ref{sec:Experimentsetup}, we assess our security method for two-vehicle platooning and later for 10-vehicle platooning under various sensor attacks, controlled by an established distributed CACC scheme \cite{ploeg2013lp}, which fulfills vehicle-following and string stability requirements in the absence of attacks. The CACC scheme is designed for a pair of platooning vehicles with guaranteed stability using the same CACC for any number of vehicles in the platoon. Performance and stability were proven experimentally in a platoon of 6 vehicles in Section~VI of \cite{ploeg2013lp}. In this paper, we demonstrate string stability, safety and comfort of this CACC under bounded attacks using our attack detection framework.

\section{Methodology}
\label{sec:methodology}
\vspace{-0.3 cm}
\subsection{Notations}

Let the real numbers be denoted by $\mathbb{R}$ ($\mathbb{R}_{> 0} = (0, \infty)$), the natural numbers be denoted by $\mathbb{N}$, the integer numbers be denoted by $\mathbb{Z}$ ($\mathbb{Z}_{\geq 0} = [0, \infty)$), and $\mathbb{R}^{n\times m}$ the set of $n\times m$ matrices with real entries for any $m,n \in \mathbb{N}$. For any vector $v\in\mathbb{R}^{n}$,  we denote  $|v|=\sqrt{v^{\top} v}$. For $v(k) \in \mathbb{R}^{n}$, time interval $
k \in \mathbb{Z}_{\geq 0}$, $||v||_{\infty} := \sup_{k\geq 0}|v(k)|$. We say that $v(k)$ belongs to $l_{\infty}$, $v(k) \in l_{\infty}$, if $||v||_{\infty} <\infty$. Similarly, $||v(k)||_{p}$ is the p-norm of signal $v(k)$.

For a set $\mathcal{N}$ let us denote by $\textrm{card}[\mathcal{N}]$ the cardinality of the set. {\small $ \left( \begin{matrix} n \\ m \end{matrix} \right)$} denotes the binomial coefficient ‘$n$ choose $m$’. $\textrm{exp}(.)$ denotes the exponential of its argument. 
An identity matrix with dimension $n \times n$, is defined as $I_n \in \mathbb{R}^{n \times n}$. $\lambda_A = [\lambda_{A,1},\dots,\lambda_{A,n}]^{\top}$ is a vector including all the eigenvalues of square matrix $A \in \mathbb{R}^{n\times n}$. The function $\textrm{floor}[.]$ takes a real number as input and gives the greatest integer less than or equal to the number as output.

\vspace{-0.3 cm}
\subsection{System dynamics}

In this paper, we focus on platoons of AVs, with longitudinal movement, in which the system dynamics of the platoons of AVs have been studied extensively by other scholars\cite{shen2022cooperative, vegamoor2021string, li2022cooperative, baldi2020establishing}. Accordingly, the majority of the previous studies considered the platooning formation as a Linear Time-Invariant (LTI) system \cite{yu2023stability, zhao2020vehicle, halder2022stability, dutta2020design}. This system may include two vehicles up to any finite number of vehicles.
Thus, to explain our problem, we investigate a generic discrete-time LTI system with a time step $T_s$, indexed by $k \in \mathbb{Z}_{\geq 0}$, where the time
is $t = T_sk$, with $p$ sensors of the form:
\begin{align}
\begin{cases}
    x(k+1) = Ax(k) + Bu(k) + w(k),  \\
    y_i(k) = C_ix(k) +  \gamma_i(k) + \delta_i(k), \hspace{2mm} i = 1,\ldots,p,  
    \end{cases}
    \label{eq:system}
\end{align}
with state $x(k) = [x_1(k),\dots,x_n(k)]^{\top} \in \mathbb{R}^{n}$, known input $u(k) \in \mathbb{R}^{m}$, system noise $w(k) \in \mathbb{R}^{n}$ with $||w(k)|| \leq \mathcal{B}_w$, $i$-th sensor measurement $y_i(k) \in \mathbb{R}$, sensor noise $\gamma_i(k) \in \mathbb{R}$ with $||\gamma_i(k)|| \leq \mathcal{B}_{\gamma}$, unknown attack signal $\delta_i(k) \in \mathbb{R}$, and known system matrices of appropriate dimensions $(A,B,C_i)$, $i = 1,\ldots,p$. Noise signals, $w(k)$ and $\gamma_i(k)$, are uniformly bounded. If  $\delta_i(k) = 0$ then the $i$-th sensor is attack-free; otherwise, sensor $i$ is under attack, and $\delta_i(k)$ is arbitrary. We assume that there are $q \in \mathbb{N}$ attacks, where $q < p$. The unknown set of attacked sensors is denoted as $\Psi \subset \{1,\ldots, p\}$, where $\textrm{card}[\Psi] = q$. 

\begin{problem} \label{prob:ratio_no_attacked}
Consider the system dynamics \eqref{eq:system} with state $x(k)$, known system matrices $(A,B,C_i)$, measured input-output trajectories $(u(k),y(k))$, and $q$ sensor attacks $\delta_i(k)$, $i \in G$, $
k \in \mathbb{Z}_{\geq 0}$. Design a state estimator $\hat{\bar{x}}(k)$ of $x(k)$ in which in the presence of bounded attack ($\delta_i, i\in\{1, \dots,p \}$), the estimation error ($e(k) = x(k) - \hat{\bar{x}}(k)$) becomes globally asymptotically stable for $\gamma_i(k) = 0$, $w(k) = \mathbf{0}$, and uniformly bounded for nonzero $\gamma_i(k)$ and $w(k)$. 
\end{problem}

\vspace{-0.3 cm}
\subsection{Generic optimal-coupling-observer framework}
\label{sec:methodology}

Based on the construction of the sensor measurements ($y_i$), we design a framework including $N \in \mathbb{N}$ number of coupled nonlinear observers for the system in \eqref{eq:system} as 
\begin{align}\begin{cases}
    \hat{x}_{J_j}(k+1) \!&=\! A \hat{x}_{J_j}(k) \!+\! Bu(k) \!+\! L_{J_j} (y_{J_j}(k) \!-\! C_{J_j}\hat{x}_{J_j}(k))\\
    &+ (1-\beta_{J_j}(k))D(\hat{\bar{x}}(k) - \hat{x}_{J_j}(k)),\\
    r_{J_j}(k) \!&=\! y_{J_j}(k) - C_{J_j}\hat{x}_{J_j}(k),
    \end{cases}
    \label{eq:observer_K}
\end{align}
where $J_j \in J$, $\text{card}[J] = N$, and $J$ denotes the set that contains designed subsets of sensors. $y_{J_j}(k) \in \mathbb{R}^{\textrm{card}[J_j]}$, $\hat{x}_{J_j}(k) \in \mathbb{R}^{n}$ is the observer state, and $r_{J_j}(k) \in \mathbb{R}^{\text{card}[J_j]}$ is a vector including the residuals of the observer. Diagonal matrix $D \in \mathbb{R}^{n\times n}$ is a weighting matrix instructed based on the eigenvalues of $A$. Nonlinear time-varying parameter $\beta_{J_j}(k) \in \mathbb{R}_{>0}$ is the classification ratio capturing the performance of each observer based on $r_{J_j}(k)$.
The term $(1-\beta_{J_j}(k))D(\hat{\bar{x}}(k) - \hat{x}_{J_j}(k))$ couples the observer $J_j$ with the most reliable output ($\hat{\bar{x}} \in \mathbb{R}^n$) belonging to the observer with the maximum classification ratio. The matrix $L_{J_j} \in \mathbb{R}^{n \times \textrm{card}[J_j]}$ is the observer gain, and $C_{J_j} \in \mathbb{R}^{ \textrm{card}[J_j] \times n} $ depends on the construction of sensors subsets.

\begin{remark}
\label{remark:eigA1}
The logic behind formulating the observer with the nonlinear time-varying term $(1-\beta_{J_j}(k))D(\hat{\bar{x}}(k) - \hat{x}_{J_j}(k))$ in \eqref{eq:observer_K} is to smoothly increase the excitability of the system in the presence of attack(s). To show the importance of embedding the term, we assume \eqref{eq:observer_K} without the nonlinear term, define the observer error as $e_{J_j}(k) = x(k)-\hat{x}_{J_j}(k)$, and introduce a perturbed version of the error dynamics, following from \eqref{eq:system}, \eqref{eq:observer_K}, as
\begin{align}
    e_{J_j}(k+1) = &(A-L_{J_j}C_{J_j}) e_{J_j}(k) \nonumber \\
    &- L_{J_j} (\gamma_{J_j}(k) + \delta_{J_j}(k)) + w(k),
    \label{eq:e_observer}
\end{align}
in which $(A-L_{J_j}C_{J_j})$ is stable (Schur stable), $\gamma_{J_j}(k) \in \mathbb{R}^{\textrm{card}[J_j]}$ includes all $\gamma_i$, $i \in J_j$, and $\delta_{J_j}(k) \in \mathbb{R}^{\textrm{card}[J_j]}$ includes all $\delta_i$, $i \in J_j$.
For simplification let $\gamma_i(k) = 0$, $w(k) = \mathbf{0}$, for any $\delta_{J_j}(k)$ that results in error convergence, namely $e_{J_j}(k+1) = e_{J_j}(k)$, from \eqref{eq:e_observer}, we have
\begin{equation}
    e_{J_j}(k) = (A-I_n-L_{J_j}C_{J_j})^{-1}L_{J_j}\delta_{J_j}(k).
    \label{eq:error_converge}
\end{equation}
From \eqref{eq:observer_K}, \eqref{eq:error_converge} and knowing $y_{J_j}(k) = C_{J_j}x(k) + \delta_{J_j}(k)$, the residual of the observer would be
\begin{align}
    \hspace{-0.1cm} r_{J_j}(k) &= C_{J_j}x(k) + \delta_{J_j}(k) - C_{J_j}\hat{x}_{J_j}(k),\nonumber \\
    &= C_{J_j}e_{J_j}(k) + \delta_{J_j}(k), \nonumber \\
    &= (C_{J_j}(A\!-\!I_n\!-\!L_{J_j}C_{J_j})^{-1}L_{J_j} \!+\! I_{\textrm{card}[J_j]})\delta_{J_j}(k).
    \label{eq:residual_ss}
\end{align}
We assume $A_{r_{J_j}} = C_{J_j}(A-L_{J_j}C_{J_j}-I_n)^{-1}L_{J_j} + I_{\textrm{card}[J_j]}$. For any eigenvalues of $A$ equal to one ($\lambda_{A,g} = 1$ corresponding to $x_g(k) \in x(k)$) and the $y_i$ that is measuring fully or partly the $x_g$, the columns of $A_{r_{J_j}}$ have only zero entries corresponding to $i \in J_{j}$. This is because the $I_n$ in $A-I_n$ bans the excitation of $r_{J_j}$ by the attacks on the sensors measure the states corresponding to $\lambda_{A,g} = 1$, known as Excitability Problem (EP). This means that the residual $r_{J_j}$ is independent of the sensor measurements $y_i$, and will not change if $y_i$ is under attack. To this end, we design the diagonal matrix $D = \textrm{diag}(D_1, \dots, D_n)$ in \eqref{eq:observer_K} such that only the entries corresponded to $\lambda_{A,g} = 1$ are equal to 1 and the rest of entries on the diagonal of $D$ are zero, i.e.,
\begin{equation}
    D_z = \begin{cases}
         1 \;\;\;\textrm{if} \; \lambda_{A,g} = 1 \;\;, \textrm{and}\; z = g, \\
        0 \;\;\;\textrm{otherwise},
    \end{cases}, z\in \{1, \dots, n\}.
    \label{eq:matrix_D}
\end{equation}
Thus, embedding a nonzero $\beta_{J_j} \in (0, 1)$, and the matrix $D$ in \eqref{eq:matrix_D} guarantees the excitability of $r_{J_j}$ by any type of bounded attacks where $D$ shifts those $\lambda_{A,g} = 1$ from 1 towards zero to solve the excitability problem.  
\color{black}
\end{remark}

\begin{example}
    To illustrate Remark~\ref{remark:eigA1}, we assume a simple system with no extended-term where $A = 1$, thus $n=1$, $\textrm{card}[J_j] = 1$, $L_{J_j} \in \mathbb{R}_{\neq 0}$, and $C_{J_j} = 1$. From \eqref{eq:residual_ss}, the residual of the observer under error convergence would be
    \begin{align}
    r_{J_j}(k) &= (C_{J_j}(1-L_{J_j}C_{J_j}-1)^{-1}L_{J_j} + 1)\delta_{J_j}(k),\nonumber  \\
    & = (-L_{J_j}^{-1}L_{J_j} + 1)\delta_{J_j}(k), \nonumber \\
    &= 0.
    \label{eq:residual_ss_example}
\end{align}
From \eqref{eq:residual_ss_example} it is shown that when the error of the observer is converging, the residuals of some observers have the EP, are converging to zero and independent of the attacks ($\delta_{J_j}$) and the observer gains $L_{J_j}$, which means that the $r_{J_j}$ is not excitable. However, if there exist the term $(1-\beta_{J_j}(k))D(\hat{\bar{x}}(k) - \hat{x}_{J_j}(k))$ in \eqref{eq:observer_K} with $\beta_{J_j} \neq 1$ for the compromised observers, the impact of $I_n$ in $A-I_n$ will be degraded and thus $r_{J_j} \neq 0$ for any $\delta_{J_j} \neq 0$.
\end{example}

\begin{figure}[t]
	\centering
	\includegraphics[width= 1\linewidth]{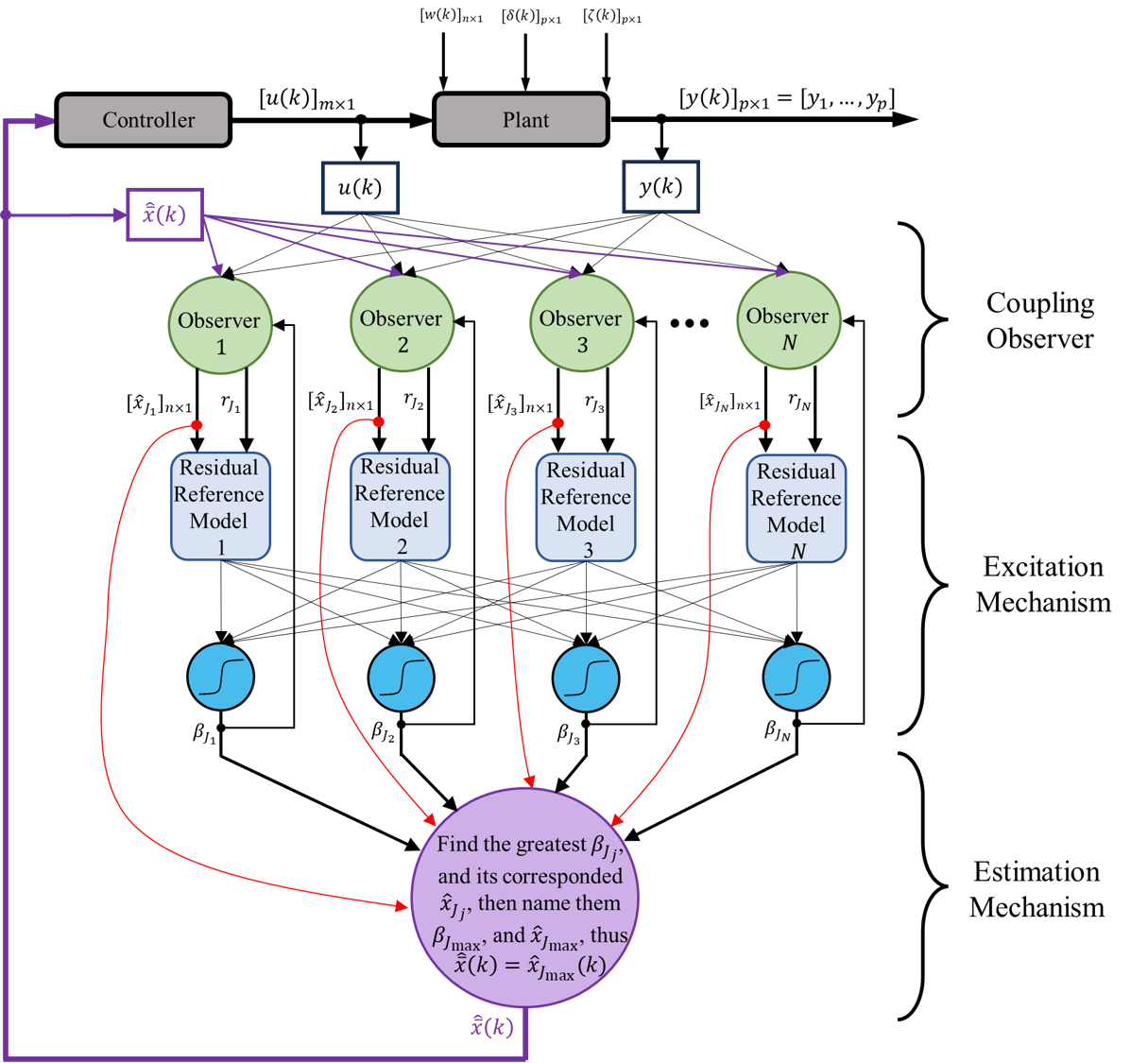}
	\caption{The optimal-coupling-observer network scheme.}
	\label{fig:plant}
    \vspace{\VDis}
\end{figure}

To practically use the suggested observer, the framework relies on designing three major components: 1) Sensor set design, 2) Excitation mechanism, and 3) Estimation mechanism. As shown in Fig.~\ref{fig:plant}, all these components together form the framework that secures the system from sensor attacks. We describe each component in the sequel.

\subsubsection{\textbf{Sensor set design}}
\label{subsec:Multi-observer}

let $J$, $\text{card}[J] = N$, $N \in \mathbb{N}$, contain all the detectable subsets of sensors where for any $J_{\rho}, J_j \in J, {\rho},j \in \{ 1, \cdots, N \}, j \neq {\rho}$, and $\textrm{card}[J_{\rho}] \leq \textrm{card}[J_j]$, we have $J_{\rho} \nsubseteq J_j$. Then we call $J_{\rho}$ and $J_j$ independent. As stated in \cite{yang2019multi}, chapter 5, the reason to skip all the dependent subsets is that any impact on the independent subset appears in the dependent subset, thus considering them only makes the computations expensive. In addition, by minimizing the $\textrm{card}[J_j]$ we increase the freedom for the variation of the number of tolerable attacks ($q$), as we know from \cite{yang2019multi} that,
\begin{equation}
    \textrm{card}[J_j] \leq q < \frac{p}{2}.
    \label{eq:num_tolerable_attack}
\end{equation}
Apparently, for any subset of sensors $J_j \in J$, the $C_{J_j} \in \mathbb{R}^{ \textrm{card}[J_j] \times n}$ that results from stacking the rows of $C$ corresponding to $y_i$, $i \in J_j$, leads to a detectable pair $(A,C_{J_j})$. 
\begin{example} Assume having a system as follows:
\begin{align}
\begin{split}
    x_1(k+1) &= x_2(k) + u_1(k), \\
    x_2(k+1) &= x_1(k) - x_2(k), 
\end{split}
\end{align}
with $p = 4$ sensors 
\begin{align}
\begin{split}
    y_i &= x_1 + \delta_i, \;\textrm{and}\;  C_i = \left[ 1 \;\;\; 0 \right], \;\;\; \textrm{for} \;\; i\in \{ 1, 2\}, \\
    y_i &= x_2 + \delta_i, \;\textrm{and}\;  C_i = \left[ 0 \;\;\; 1 \right], \;\;\; \textrm{for} \;\; i\in \{ 3, 4 \}. 
\end{split}
\label{eq:example1}
\end{align}
First, note that this system is observable in the usual sense considering $A = ${\small $\left[ \begin{matrix} 0 & 1 \\ 1 & -1 \end{matrix} \right]$}. Accordingly, the number of all possible sensor combinations is {\small $ \sum_{s=1}^{4}\left( \begin{matrix} 4 \\ s \end{matrix} \right) = 15$}. By checking Hautus lemma\cite{sontag2013mathematical} for detectability check for each subset, we find out that all 15 subsets are detectable. Considering the detectable set $\Phi = \{\Phi_1, \cdots, \Phi_{15}\}$, where $\Phi_1 = \{ 1 \}$, $\Phi_2 = \{ 2 \}$, $\Phi_3 = \{ 3 \}$, $\Phi_4 = \{ 4 \}$, $\Phi_5 = \{ 1,2 \}$, $\Phi_6 = \{ 1,3 \}$, $\Phi_7 = \{ 1,4 \}$, $\Phi_8 = \{ 2,3 \}$, $\Phi_9 = \{ 2,4 \}$, $\Phi_{10} = \{ 3,4 \}$, $\Phi_{11} = \{ 1,2,3 \}$, $\Phi_{12} = \{ 1,2,4 \}$, $\Phi_{13} = \{ 1,3,4 \}$, $\Phi_{14} = \{ 2,3,4 \}$, and $\Phi_{15} = \{ 1,2,3,4 \}$. As, $J$ should contain the independent subsets of $\Phi$, thus $J=\{ \Phi_1, \Phi_2, \Phi_3, \Phi_4 \}$, $\textrm{card}[J] = 4$. Accordingly, $J_1 = \Phi_1$, $J_2 = \Phi_2$, $J_3 = \Phi_3$, and $J_4 = \Phi_4$, and $\textrm{card}[J_j] = 1$, $j \in \{1, \dots, 4 \}$, thus from \eqref{eq:num_tolerable_attack}, the system is one attack tolerable ($q = 1$).
\end{example}

\subsubsection{\textbf{Excitation mechanism}}
\label{subsec:excitation_mechanism}

Using the $N$ observers, our purpose is to classify the observers based on their degree of vulnerability against accrued attacks. In the best-case scenario, the criterion used to evaluate the performance of each observer would depend on the observer error $e_{J_j}(k) = x(k) - \hat{x}_{J_j}(k)$ for ${J_j} \in J$. However, the state $x(k)$ is unknown which makes $e_{J_j}(k)$ become unknown, and any performance criterion involving $e_{J_j}(k)$ would not be implementable. As a result and similarly to other works e.g.,~\cite{willems2004deterministic, na2017adaptive, chong2015parameter}, we use the knowledge from the sensors sets $y_{J_j}(k)$, and the corresponding observer states $\hat{x}_{J_j}(k)$ with $j \in \{1, \cdots, N \}$. The excitation mechanism includes the following components:

\paragraph{\underline{Residual reference model}}
we introduce a residual reference model ($\eta_{J_j}$) excited by the residual of each observer employing the well-known mass-spring-damper model~\cite{armaghan2011design} with mass value 1, which is a two-state system with globally asymptotically stable dynamics around $|r_{J_j}|$ as follows.
\begin{equation}
    x_{r_{J_j}}(k+1) = A_r x_{r_{J_j}}(k) + B_r |r_{J_j}(k)|,
\end{equation}
where
\begin{equation}
\begin{cases}
    A_r = \textrm{exp}(A_r^c T_s), B_r = \stretchint{5ex}_{\!\! 0}^{T_s} \textrm{exp}\left(A_r^c (T_s - s)\right) B_r^c ds,
    \\
    A_r^c = \left[\begin{matrix}
        0 & 1 \\
        -K_r & -C_r
    \end{matrix}\right], 
    B_r^c = \left[\begin{matrix}
        0 \\
        K_r 
    \end{matrix}\right], 
    \end{cases}
    \label{eq:matrices_dynamic_dis}
\end{equation}
$x_{r_{J_j}} := \left[ \begin{matrix} x_{r_{J_j}}^{1,1} & x_{r_{J_j}}^{2,1} \end{matrix} \right]^{\top}$ and $\eta_{J_j}(k) = x_{r_{J_j}}^{1,1}(k)$. The system is globally asymptotically stable if all the eigenvalues of $A_{r}^c$ are negative and the pair ($A_{r}^c, B_{r}^c$) is stabilisable (see, e.g.,~\cite{ref:Williams2007}).

\paragraph{\underline{Classification ratio}}
\label{subsec:Scoring_ratio}

To give a logical comparison between the observers, we design a classifier term $\beta^{\eta}_{J_j}(k)  \in \mathbb{R}$, $\beta^{\eta}_{J_j}(k) \in (0,1)$ with ${J_j} \in J$ for each of the observers which is dependant to all the $\eta_{J_j}(k)$ ${J_j} \in J$, that contains higher values for reliable sets comparing to the compromised sets (see Remark~\ref{remark:healthy_attacked}). To this end, we design the classifier term for the $j^{\textrm{th}}$ observer as below:
\begin{align}
    \beta^{\eta}_{J_j}(k) = 1-\frac{\eta_{J_j}(k)+\mathcal{B}_w+\mathcal{B}_{\gamma}}{\sum^{N}_{s = 1} \left(\eta_{J_s}(k)+\mathcal{B}_w+\mathcal{B}_{\gamma}\right)},\;\;\; \beta_{J_j}(0) \in (0,1).
\label{eq:beta}
\end{align}
According to \eqref{eq:beta}, we define $\bar{\beta}^{\eta} = 1-\frac{1}{N}$ as the classifier term value for the safe condition where $q=0$ and thus $\eta_{J_j}$ converges to zero for any $J_j \in J$. 

\begin{remark}
    If $J_j \cap \Psi = \varnothing$, and $\Psi \neq \varnothing$, subset $J_j \in J$ is a reliable set and the corresponding $\eta_{J_j}$ convergence to 0, and $\beta_{J_j}$ converges to 1; otherwise, $J_j$ is a compromised subset and consequently $\eta_{J_j} \leq \bar{\eta}_{J_j}$ for some non-negative constant $\bar{\eta}_{J_j}$.
    \label{remark:healthy_attacked}
\end{remark}

Although the introduced classifier term in \eqref{eq:beta} offers a higher value close to 1 for the reliable set(s), the term is least likely to offer a value close to 0, especially when $q\geq2$. To make the mechanism more sensitive to reliable and compromised sets, we use the following activation function to design the classification ratio.
\begin{equation}
    \beta_{J_j}(k) = \frac{1}{\pi}\textrm{arctan}\left((\beta^{\eta}_{J_j}(k) - \bar{\beta}^{\eta})a_{\beta}\right)+0.5,
    \label{eq:classification_ratio}
\end{equation}
where, $\beta_{J_j}(k) \in \mathbb{R}$, $\beta_{J_j}(k) \in (0,1)$ is the classification ratio, and $a_{\beta}$ is the magnifier parameter which regulates the sensitivity of the classification ratio to attacks. Accordingly, if $q=0 \; (\Psi = \varnothing)$, then $\beta^{\eta}_{J_j}(k) = \bar{\beta}^{\eta}$, and thus $\beta_{J_j}(k) = 0.5$. 

\subsubsection{\textbf{Estimation mechanism}}
Given the $N$ observers' outputs, and the $N$ classification ratios, we present an estimation mechanism that estimates the unknown state $x(k)$, such that 
\begin{equation}
    \hat{\bar{x}}(k) = \hat{x}_{J_{j}}(k), \;\; \beta_{J_j}(k) = \beta_{J_{\textrm{max}}}(k) , 
    \label{eq:estimated_state}
\end{equation}
where the estimated state $\hat{\bar{x}}(k) \in \mathbb{R}^{n}$, $\beta_{J_{\textrm{max}}}(k)$ is the maximum of all the $\beta_{J_j}(k)$, and $\hat{x}_{J_{j}}(k)$ belongs to the observer with the classification ratios value equal to the best achievable $\beta_{J_{\textrm{max}}}(k)$ which is supposed to be the most reliable observer. In the case that there are several observers with reliable sets, i.e., several $\beta_{J_j}(k)$ are equal to $\beta_{J_{\textrm{max}}}(k)$, we determine $\hat{\bar{x}}(k)$  equal to the reliable observer that has the lowest $j$.

\vspace{-0.3 cm}
\subsection{Stability Discussion}
\label{sec:Stability Discussion}

\subsubsection{\textbf{Error dynamics system}}
We define the estimation error using \eqref{eq:system}, and \eqref{eq:estimated_state} as follows
\begin{align}
    e(k) &= x(k) - \hat{\bar{x}}(k),\\
    e(k+1) &= (A-L_{J_{\textrm{max}}}(k)C_{J_{\textrm{max}}}(k)) e(k) \nonumber \\
    &- L_{J_{\textrm{max}}}(k) (\gamma_{J_{\textrm{max}}}(k) + \delta_{J_{\textrm{max}}}(k)) + w(k). 
    \label{eq:sum_error}
\end{align}
We also define the observer error dynamics following from \eqref{eq:system} and \eqref{eq:observer_K}, and knowing $-(1-\beta_{J_j}(k))D(\hat{\bar{x}}(k) - \hat{x}_{J_j}(k)) = (1-\beta_{J_j}(k))De(k) - (1-\beta_{J_j}(k))De_{J_j}(k)$, as
\begin{align}
    e_{J_j}&(k+1) = (A-(1-\beta_{J_j}(k)) D-L_{J_j}C_{J_j}) e_{J_j}(k) \nonumber \\
    &- L_{J_j} (\gamma_{J_j}(k) + \delta_{J_j}(k)) + w(k) + (1-\beta_{J_j}(k))D e(k).
    \label{eq:e_observer_K}
\end{align}
Considering \eqref{eq:sum_error} and \eqref{eq:e_observer_K}, the overall error system dynamics for the observers $J_j$ and the estimation error is derived as follows
\begin{align}
    &E_{J_j}(k+1) \!=\! \mathbb{A}_{J_j}(k)E_{J_j}(k) \!-\! \mathbb{L}_{J_j}(\Gamma_{J_j}(k) \!+\! \Delta_{J_j}(k)) \!+\! W(k), \nonumber
    \\
    &E_{J_j}(k) \!=\!\! \left[\begin{matrix}
        e_{J_j}(k) \\ e(k)
    \end{matrix}\right]\!, 
    \Gamma_{J_j}(k) \!=\!\! \left[\begin{matrix}
        \gamma_{J_j}(k) \\ \gamma_{J_{\textrm{max}}}(k)
    \end{matrix}\right]\!,
    \Delta_{J_j}(k) \!=\!\! \left[\begin{matrix}
        \delta_{J_j}(k) \\ \delta_{J_{\textrm{max}}}(k)
    \end{matrix}\right], \nonumber
    \\
    &\mathbb{A}_{J_j}(k) \!=\! \left[\begin{matrix}
        A\!-\!(1\!-\!\beta_{J_j}(k)) D\!-\!L_{J_j}C_{J_j}\!\!\!& \!\!\!(1-\beta_{J_j}(k)) D \\
        0 \!\!\!& \!\!\!\!\!\! A\!-\!L_{J_{\textrm{max}}}(k)C_{J_{\textrm{max}}}(k)
    \end{matrix}\right], \nonumber
    \\
    &\mathbb{L}_{J_j}(k) = \left[\begin{matrix}
        L_{J_j} & 0 
        \\
        0 & L_{J_{\textrm{max}}}(k)
    \end{matrix}\right], W(k) = \left[\begin{matrix}
        w(k) \\ w(k)
    \end{matrix}\right],
    \label{eq:completeErrorDynamic}
\end{align}
where $E_{J_j}(k) \in \mathbb{R}^{2n}$, $\Gamma_{J_j}(k) \in \mathbb{R}^{\textrm{card}[J_j] + \textrm{card}[J_{\textrm{max}}]}$, $\Delta_{J_j}(k) \in \mathbb{R}^{\textrm{card}[J_j] + \textrm{card}[J_{\textrm{max}}]}$, $W(k) \in \mathbb{R}^{2n}$, and $\mathbb{L}_{J_j} \in \mathbb{R}^{2n \times ({\textrm{card}[J_j] + \textrm{card}[J_{\textrm{max}}]})}$. 

\subsubsection{\textbf{Stability discussion and parameters design}}

Regarding the error dynamics system explained in \eqref{eq:completeErrorDynamic}, here, the purpose is to design matrices $L_{J_j}$ such that the complete error dynamic system \eqref{eq:completeErrorDynamic} becomes asymptotically stable in the absence of attack and noise. Due to the construction of matrix $\mathbb{A}_{J_j}$ which is an upper triangle matrix in \eqref{eq:completeErrorDynamic}, complete error dynamic after $k$ time intervals in the absence of attack, and noise ($\Gamma_{J_j} = \mathbf{0}, \Delta_{J_j} = \mathbf{0}, W = \mathbf{0}$) is 
\begin{align}
E_{J_j}(k+1) = \left(\prod_{\iota=0}^{k}\mathbb{A}_{J_j}(\iota)\right) E_{J_j}(0), 
\label{eq:completeErrSys}
\end{align}
where
\begin{align}
    & \prod_{\iota=0}^{k}\mathbb{A}_{J_j}(\iota) = \left[\begin{matrix}
        \mathbb{A}_{1,1}^{\prod}(k) & \star \\
        \mathbf{0} & \mathbb{A}_{2,2}^{\prod}(k) 
    \end{matrix}\right], \nonumber 
    \\
    &\begin{cases}
        \mathbb{A}_{1,1}^{\prod}(k) = \prod_{\iota=0}^{k} A-(1-\beta_{J_j}(\iota)) D-L_{J_j}C_{J_j}, \\
        \mathbb{A}_{2,2}^{\prod}(k) = \prod_{\iota=0}^{k} A-L_{J_{\textrm{max}}}(\iota)C_{J_{\textrm{max}}}(\iota).
    \end{cases}
    \label{eq:completeA}
\end{align}
As $\prod_{\iota=0}^{k}\mathbb{A}_{J_j}(\iota)$ is an upper triangle matrix, the complete error system is asymptotically stable if and only if the matrices on the diagonal of $\prod_{\iota=0}^{k}\mathbb{A}_{J_j}(\iota)$ converges to 0 when $k \rightarrow \infty$.  To satisfy this condition, we will design $L_{J_j}$ by using the LMI method.

\begin{remark}
\label{remark:linearalgebra}
In \eqref{eq:completeA}, there are two time-varying matrices on the diagonal of $\prod_{\iota=0}^{k}\mathbb{A}_{J_j}(\iota)$ as $\mathbb{A}_{1,1}^{\prod}(k)$ and $\mathbb{A}_{2,2}^{\prod}(k)$. Note that $\mathbb{A}_{1,1}^{\prod}(k)$ is a stronger time-varying matrix than $\mathbb{A}_{2,2}^{\prod}(k)$ as it includes a time-varying component of $\beta_{J_j}$, and asymptotic stability of $\mathbb{A}_{1,1}^{\prod}(k)$ results in the asymptotic stability of $\mathbb{A}_{2,2}^{\prod}(k)$ as $\mathbb{A}_{2,2}^{\prod}(k)$ is equal to $\mathbb{A}_{1,1}^{\prod}(k)$ when $\beta_{J_j}(k) = 1$. Thus, designing $L_{J_j}$ for asymptotic stability of $\mathbb{A}_{1,1}^{\prod}(k)$ for the extreme conditions of $\beta_{J_j}$ (0 and 1) is enough to guarantee the asymptotic stability of the complete error system in \eqref{eq:completeErrSys}.
\end{remark}

First, we explain the inequalities in the form of discrete Lyapunov functions that can be reformulated as an LMI problem using the Schur complement approach \cite{li2015design}. Regarding Remark~\ref{remark:linearalgebra}, the equivalent inequality to be solved by the LMI method for the condition that $\beta_{J_j} = 1$ is:
\begin{equation}
    (A-L_{J_j}C_{J_j})^{\top}P(A-L_{J_j}C_{J_j}) - P < 0, \;\; J_j \in J,
    \label{eq:LMI:A-LC}
\end{equation}
where $P\in \mathbb{R}^{n \times n}_{>0}$ is a positive definite matrix. 
Similarly, the equivalent inequality to be solved by the LMI method for the condition that $\beta_{J_j} = 0$ is:
\begin{equation}
    (A-D-L_{J_j}C_{J_j})^{\top}P(A-D-L_{J_j}C_{J_j}) - P < 0, \;\; J_j \in J.
    \label{eq:LMI:A-K-LC}
\end{equation}
Accordingly, the LMI problem includes $2N$ inequalities in \eqref{eq:LMI:A-LC}-\eqref{eq:LMI:A-K-LC}, and one inequality for $P>0$ that should be satisfied by designing a unique $P$ and proper $L_{J_j}$ with $ j\in \{1, \cdots, N \}$.

Using the Schur complement method \cite{li2015design}, the $N$ inequalities in \eqref{eq:LMI:A-LC} can be turned into an equivalent LMI problem as
\begin{equation}
    \left[\begin{matrix}
        -P & P^{\top}A-Z_{J_j}^{\top}C_{J_j}\\
        A^{\top}P-C_{J_j}^{\top}Z_{J_j} & -P
    \end{matrix}\right] < 0,
    \label{eq:lmi_A_P}
\end{equation}
the extra $N$ inequalities in \eqref{eq:LMI:A-K-LC} can be turned into an equivalent LMI problem as
\begin{equation}
    \left[\begin{matrix}
        -P & P^{\top}(A-D)-Z_{J_j}^{\top}C_{J_j}\\
        (A-D)^{\top}P-C_{J_j}^{\top}Z_{J_j} & -P
    \end{matrix}\right] < 0,
    \label{eq:lmi_A_In_P}
\end{equation}
while,
\begin{equation}
    P > 0.
    \label{eq:lmi_P}
\end{equation}
We solved all the $2N+1$ inequalities in \eqref{eq:lmi_A_P}-\eqref{eq:lmi_P} to find $P$, and $Z_{J_j}$, $J_j \in J$, where
\begin{equation}
    L_{J_j} = \left( Z_{J_j} P^{-1} \right)^{\top}.
    \label{eq:L_Jj_Design}
\end{equation}

\begin{remark}
\label{remark:LMI_OffLine}
The LMI problem including the $2N + 1$ inequalities in \eqref{eq:lmi_A_P}-\eqref{eq:lmi_P} is solved off-line, and thus the value of $N$ and solving time have no impact on the real-time implementation of the method in practice.
\end{remark}

\begin{remark}
\label{remark:LMI_Solvability}
The LMI problem including the $2N + 1$ inequalities in \eqref{eq:lmi_A_P}-\eqref{eq:lmi_P} should have a unique solution as long as each pair ($A, C_{J_j}$), and ($A-D, C_{J_j}$) is detectable. The reason is that the detectability of each pair ($A, C_{J_j}$) defines the feasibility of \eqref{eq:lmi_A_P}, and the detectability of each pair ($A-D, C_{J_j}$) defines the feasibility of \eqref{eq:lmi_A_In_P}. Due to the instruction of $D$ in \eqref{eq:matrix_D}, pair ($A-D, C_{J_j}$) gives equal detectability to  pair ($A, C_{J_j}$) (if none of the eigenvalues of $A$ is equal to 1) or stronger detectability than ($A, C_{J_j}$) (if at least one of the eigenvalues of $A$ is equal to 1). And as $C_{J_j}$ is designed to guarantee detectability of pair ($A, C_{J_j}$), thus ($A-D, C_{J_j}$) should be detectable in general sense. Note that matrix $D$ does not make the matrix $A-D$ Schur stable, and it only improves the detectability of pair ($A-D, C_{J_j}$) compared to pair ($A, C_{J_j}$).

\end{remark}
\color{black}

\vspace{-0.1 cm}
\section{Simulation Setup}
\label{sec:Experimentsetup}
\vspace{-0.1 cm}
\subsection{Vehicle platooning model}
\label{subsec:Vehicle_platooning}
\begin{figure}[tb]
	\centering
	\includegraphics[width= 1\linewidth]{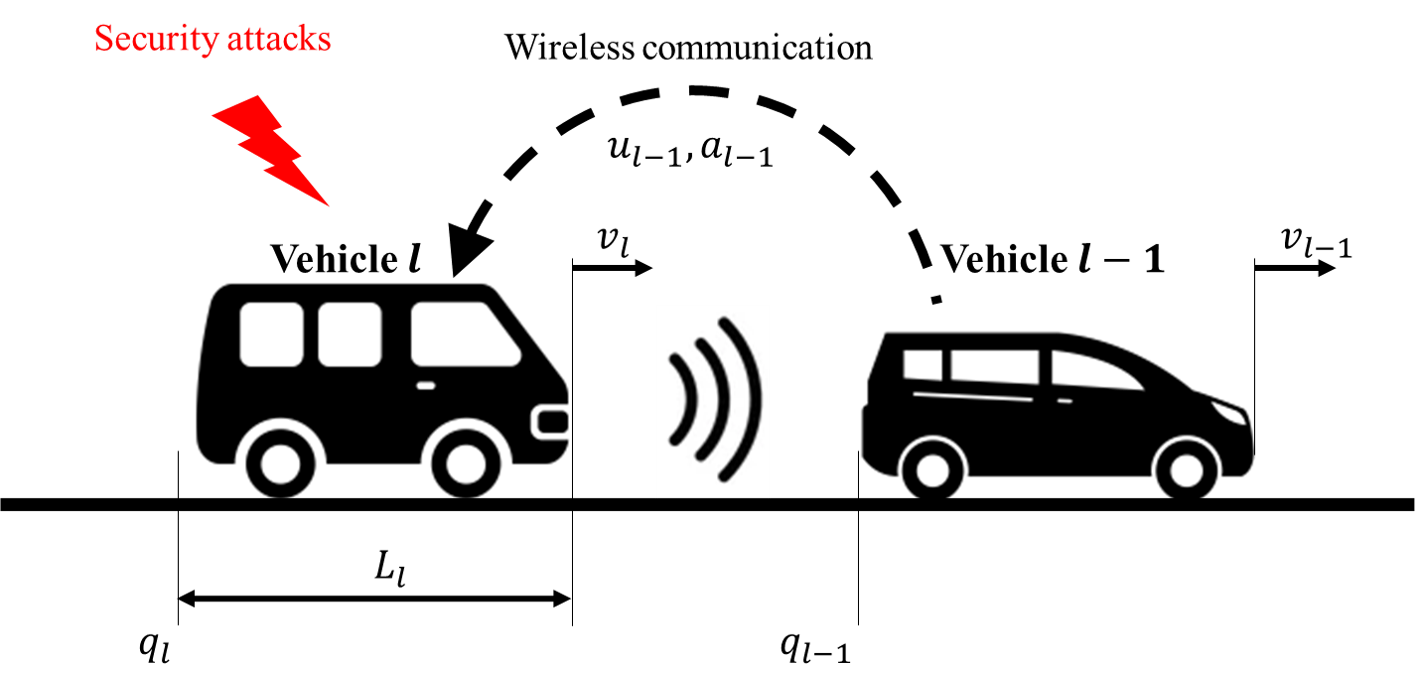}
	\caption{CACC-equipped vehicles.}
	\label{fig:ExampleModel}
    \vspace{\VDis}
\end{figure}

Similar to \cite{yang2021secure}, we consider a string of $M \in \mathbb{N}$ platooning vehicles, schematically depicted in Fig.~\ref{fig:ExampleModel}, with $d_l$ being the distance between vehicle $l$ and its preceding vehicle $l-1$, and $v_l$ the velocity of vehicle $l$.
The main objective of each vehicle is to follow its preceding (lead) vehicle at a desired distance $d_{r,l}$. Here, a constant time headway spacing policy is adopted, formulated as
\begin{equation}
    d_{r,l}(t) = s_l+hv_l(t), l\in S_M,
    \label{eq:dri}
\end{equation}
where $h>0$ is the so-called time headway, $s_l$ is the standstill distance, and $S_M:=\{ l \in \mathbb{N} | l \leq M  \}$. This spacing policy is known to improve string stability \cite{naus2010string, rajamani2002semi, klinge2009string}. A homogeneous string is assumed,
which is why the time headway $h$ is taken independently of $l$. The spacing error $e_l(t)$ is thus defined as
\begin{align}
    e_l(t) &= d_l(t) - d_{r,l}(t), \nonumber \\
    &= \left( q_{l-1}(t) - q_l(t) - \mathcal{L}_l \right) - \left( s_l + h v_l(t) \right),
    \label{eq:e_i}
\end{align}
with $q_l(t)$ is the position of vehicle $l$ and $\mathcal{L}_l$ is its length, and $d_l(t) = q_{l-1}(t) - q_l(t) - \mathcal{L}_l$ is the distance between the vehicle $l$ and its preceding vehicle $l-1$. As a basis for control design, the following vehicle model is adopted from \cite{ploeg2013lp}
\begin{equation}
    \left[ \begin{matrix}
        \dot{d}_l(t) \\ \dot{v}_l(t) \\ \dot{a}_l(t)
    \end{matrix} \right] = 
    \left[ \begin{matrix}
        v_{l-1}(t) - v_l(t)\\ a_l(t) \\ -\frac{1}{\tau} a_l(t) + \frac{1}{\tau} u_l(t)
    \end{matrix} \right], l \in S_M,
    \label{eq:modelExp}
\end{equation}
where $\tau$ is a time constant modelling driveline dynamics (namely, $\tau = 0.1$ s in \cite{ploeg2013lp}), $a_l$ denotes the acceleration of vehicle $l$, and $u_l(t)$ is its desired acceleration (the control input). 

\vspace{-0.3 cm}
\subsection{Vehicle platooning control}
To enforce the platooning behavior (by designing controller $u_l$ in  \eqref{eq:modelExp}), we exactly adopt the distributed CACC scheme introduced in \cite{ploeg2013lp}, which fulfills the vehicle-following objective and string stability requirement in the absence of attacks \cite{ploeg2013lp}. This controller $u_l(t)$ is dynamic and can be written as follows: 
\begin{align}
\begin{cases}
    \xi_l(t) = k_p e_l(t) + k_d \dot{e}_l(t) + k_{dd} \ddot{e}_l(t) + u_{l-1}(t),
    \\
    \dot{u}_l(t) = -\frac{1}{h}u_l(t) + \frac{1}{h}\xi_l(t),
    \end{cases}
    \label{eq:cont111}
\end{align}
where $l \in S_m$, $\xi_l \in \mathbb{R}$ is the controller state and constants $k_p \in \mathbb{R}$, $k_d \in \mathbb{R}$, and $k_{dd} \in \mathbb{R}$ are the control gains, i.e., to have asymptotic stability according to \cite{ploeg2013lp} it is sufficient if $k_p, k_d > 0$, $k_{dd} > -1$ and $(1+k_{dd})k_d > \tau k_p$. The feedforward term $u_{l-1}$ is the desired acceleration of the preceding vehicle obtained through wireless communication.

\vspace{-0.3 cm}
\subsection{Communication and sensing topology}
We adopt a decentralized communication topology, where vehicles are interconnected by a set of wireless communication links that allow each vehicle $l-1$ (lead vehicle) to send its acceleration to vehicle $l$ (following vehicle). For control and monitoring purposes, we further use sensor data coming from onboard sensors (e.g., radar, LiDAR, cameras, sensing distance and relative velocity to the lead vehicle). Please note that, as indicated in \eqref{eq:sensors9}, when measuring the same variables with different sensors, we utilize three sensors for the distance ($d_l(t)$), specifically $y_1(t)$, $y_6(t)$, and $y_8(t)$, as well as three sensors for velocity, namely $y_2(t)$, $y_7(t)$, and $y_9(t)$. Without loss of generality, we assume that available sensors for vehicle $l$ are:
\vspace{-0.2 cm}
\begin{subequations}
\label{eq:sensors9}
\begin{align}
    \text{\textrm{Variable}} \; & \hspace{1cm} \text{\textrm{Related sensor measures the variable}} \notag \\
    \hline  
    d_l(t) \; &\quad \!y_1(t) \!=\! q_{l-1}(t) \!-\! q_l(t) \!-\! \mathcal{L}_l \!+\! \gamma_1(t) \!+\! \delta_1(t), \\
    v_l(t) \; &\quad \!y_2(t) \!=\! v_l(t) + \gamma_2(t) + \delta_2(t), \\
    a_l(t) \; & \quad \!y_3(t) \!=\! a_l(t) + \gamma_3(t) + \delta_3(t), \\
    \Delta v_l(t) \; & \quad \!y_4(t) \!=\! v_{l-1}(t) - v_l(t) + \gamma_4(t) + \delta_4(t), \\
    a_{l-1}(t) \; & \quad \!y_5(t) \!=\! a_{l-1}(t) + \gamma_5(t) + \delta_5(t), \\
    d_l(t) \; & \quad \!y_6(t) \!=\! q_{l-1}(t) \!-\! q_l(t) \!-\! \mathcal{L}_l \!+\! \gamma_6(t) \!+\! \delta_6(t), \\
    v_l(t) \; & \quad \!y_7(t) \!=\! v_l(t) + \gamma_7(t) + \delta_7(t), \\
    d_l(t) \; & \quad \!y_8(t) \!=\! q_{l-1}(t) \!-\! q_l(t) \!-\! \mathcal{L}_l \!+\! \gamma_8(t) \!+\! \delta_8(t), \\
    v_l(t) \; \raisebox{0pt}[0pt][0pt]{\rule{0.4pt}{9.5\baselineskip}} & \quad \!y_9(t) \!=\! v_l(t) + \gamma_9(t) + \delta_9(t),
\end{align}
\end{subequations}
where $y_{s}(t)$ denotes sensor measurements, $\delta_{s}(t)$ models potential FDI attacks that tamper with hardware, networks, and computers, and $\gamma_{s}(t)$ denotes reliable (normal) sensor noise with $s \in \{1, \dots, 9\}$. In addition, $\Delta v_l$ defines the relative velocity between vehicle $l$ and its preceding vehicle (vehicle $l-1$). 
Note that sensors $y_{1}(t)$, $y_{6}(t)$, $y_{8}(t)$ and $y_{4}(t)$ provide relative tracking and relative velocity information, and sensors $y_{2}(t)$, $y_{7}(t)$, $y_{9}(t)$ and $y_{3}(t)$ model onboard measured velocity and acceleration. In addition, $y_{5}(t)$ models acceleration data received wirelessly from the lead vehicle. Moreover, we know from \eqref{eq:sensors9} that $\dot{e}(t) = v_{l-1}(t) - v_{l}(t) - ha_{l}(t)$ measurable by $y_{4}$ and $y_{3}$. Similarly, $\ddot{e}(t) = a_{l-1}(t) - a_{l}(t) - h\dot{a}_{l}(t) = a_{l-1}(t) + (\frac{h}{\tau} - 1) a_{l}(t) - \frac{h}{\tau} u_l(t)$ measurable by $y_{3}$ and $y_{5}$. Thus, $y_1, \dots, y_9$ in \eqref{eq:sensors9} are the sensor inputs to our framework, resulting in $\hat{\bar{x}}$ in the estimation mechanism \eqref{eq:estimated_state} which will be used to define the control signal \eqref{eq:cont111} as follows.

\vspace{-0.3 cm}
\subsection{Continuous-time closed-loop dynamics}
Next, consider controller \eqref{eq:cont111} and let $y \in \mathbb{R}^9$ denote the vector of stacked sensor, i.e.,  $y := \textrm{col}[y_{1}, \dots , y_{9}]^{\top}$ , $\delta \in \mathbb{R}^9$ denote the vector of stacked sensor attacks, i.e., $\delta:= \textrm{col}[\delta_{1}, \dots , \delta_{9}]^{\top}$, and $\gamma \in \mathbb{R}^9$ denote the vector of stacked sensor noise, i.e., $\gamma:= \textrm{col}[\gamma_{1}, \dots , \gamma_{9}]^{\top}$. Define the stacked state vector $x_l:= \textrm{col}[e_l, v_l, a_l, \Delta v_l, a_{l-1}]^{\top}$, where $\Delta v_l := v_{l-1} - v_l$ is the relative velocity between vehicle $l$ and its preceding vehicle (vehicle $l-1$), carries information from $u_{l-1}$ as $\Delta \dot{v}_l = a_{l-1} - a_{l}$. In addition, similar to \eqref{eq:modelExp}, $\dot{a}_{l-1} = -\frac{1}{\tau} a_{l-1}(t) + \frac{1}{\tau} u_{l-1}(t)$, where $u_{l-1}(t)$ is assumed to be securely available. Using this notation and considering that the $\hat{\bar{x}}:= \textrm{col}[\hat{\bar{x}}_1, \hat{\bar{x}}_2, \hat{\bar{x}}_3, \hat{\bar{x}}_4, \hat{\bar{x}}_5]^{\top}$ in \eqref{eq:estimated_state} from our framework is the estimation of $x_l$, the closed-loop dynamics \eqref{eq:modelExp}-\eqref{eq:cont111} can be written as follows:
\begin{align}
\begin{cases}
    \dot{x}_l(t) &= A^{c} x_l(t) + B_1^{c} u_{l-1}(t) + B_2^{c} u_{l}(t), 
    \\
    \xi_l(t) &= k_p \hat{\bar{x}}_1(t) + k_d (\hat{\bar{x}}_{4}(t) - h\hat{\bar{x}}_{3}(t)) 
    \\
    &+ k_{dd}(\hat{\bar{x}}_{5}(t) + (\frac{h}{\tau} - 1) \hat{\bar{x}}_{3}(t) - \frac{h}{\tau} u_l(t)) + u_{l-1}(t),\\
    \dot{u}_l(t) &= -\frac{1}{h}u_l(t) + \frac{1}{h} \xi_l(t), \\
    y(t) &= C x_l(t) + \gamma(t) + \delta(t),
    \end{cases}
    \label{eq:dynamicSysExample}
\end{align}
where
\begin{align}
\begin{cases}
    A^c \!\!\!\!\!\!&= \left[ \begin{matrix}
        0 & 0 & -h & 1 & 0 \\
        0 & 0 &  1 & 0 & 0 \\
        0 & 0 & -\frac{1}{\tau} & 0 & 0\\
        0 & 0 & -1 & 0 & 1 \\
        0 & 0 & 0 & 0 & -\frac{1}{\tau}
    \end{matrix} \right],
    B_1^c = \left[ \begin{matrix}
        0 \\ 0 \\ 0 \\ 0 \\ \frac{1}{\tau}
    \end{matrix} \right], \\
    B_2^c &= \left[ \begin{matrix}
        0 \\ 0 \\ \frac{1}{\tau} \\ 0 \\ 0
    \end{matrix} \right], 
    C = \left[ \begin{matrix}
        1 & 0 & 0 & 0 & 0 \\
        0 & 1 & 0 & 0 & 0 \\
        0 & 0 & 1 & 0 & 0 \\
        0 & 0 & 0 & 1 & 0 \\
        0 & 0 & 0 & 0 & 1 \\
        1 & 0 & 0 & 0 & 0 \\
        0 & 1 & 0 & 0 & 0 \\
        1 & 0 & 0 & 0 & 0 \\
        0 & 1 & 0 & 0 & 0 
    \end{matrix} \right].
    \end{cases}
    \label{eq:matrices_dynamic}
\end{align}

\vspace{-0.3 cm}
\subsection{Discrete-time closed-loop dynamic}
Due to the fact that the controller implementation in practice is feasible in discrete time, we exactly discretized \eqref{eq:dynamicSysExample} at the sampling time instants, $t=T_sk, k\in\mathbb{Z}_{\geq0}$, assuming a zero-order hold to implement control actions and model discrete-time uncertainties and obtain the equivalent discrete-time systems:
\begin{align}
\begin{cases}
    x_l(k+1) &= A x_l(k) + B_1 u_{l-1}(k) + B_2 u_{l}(k), 
    \\
    \xi_l(k) &= k_p \hat{\bar{x}}_1(k) + k_d (\hat{\bar{x}}_{4}(k) - h\hat{\bar{x}}_{3}(k)) 
    \\
    &+ k_{dd}(\hat{\bar{x}}_{5}(k) \!+\! (\frac{h}{\tau} \!-\! 1) \hat{\bar{x}}_{3}(k) \!-\! \frac{h}{\tau} u_l(k)) \!+\! u_{l-1}(k),\\
    u_l(k+1) &= \textrm{exp}(-\frac{T_s}{h})u_l(k) + (1-\textrm{exp}(-\frac{T_s}{h})) \xi_l(k), \\
    y(k) &= C x_l(k) + \delta(k) + \gamma(k),
    \end{cases}
    \label{eq:dynamicSysExample_dis}
\end{align}
with $x_l(k):=x_l(T_sk)$, $u_{l-1}(k):=u_{l-1}(T_sk)$, $u_{l}(k):=u_{l}(T_sk)$, $\xi_l(k):=\xi_l(T_sk)$, $\hat{\bar{x}}(k):=\hat{\bar{x}}(T_sk)$, $y(k):=y(T_sk)$, $\delta(k):=\delta(T_sk)$, $\gamma(k):=\gamma(T_sk)$, and matrices
\begin{equation}
\begin{cases}
    A = \textrm{exp}(A^c T_s), B_1 = \stretchint{5ex}_{\!\! 0}^{T_s} \textrm{exp}\left(A^c (T_s - s)\right) B^c_1 ds,
    \\
    B_2 = \stretchint{5ex}_{\!\! 0}^{T_s} \textrm{exp}\left(A^c (T_s - s)\right) B^c_2 ds. 
    \end{cases}
    \label{eq:matrices_dynamic_dis_controller}
\end{equation}
The discrete-time platooning system described in \eqref{eq:dynamicSysExample_dis} constitutes a specific instance of the general linear time-invariant (LTI) system structure given in \eqref{eq:system}. In both formulations, the system matrix \( A \) retains an identical mathematical form. The input matrix \( B \) and the input vector \( u \) in \eqref{eq:system} correspond, respectively, to 
\(
\begin{bmatrix}
    B_1 & B_2
\end{bmatrix}
\) 
and 
\(
\begin{bmatrix}
    u_{l-1} & u_l
\end{bmatrix}^{\top}
\),
as derived from the structure of \eqref{eq:dynamicSysExample_dis}.

\vspace{-0.3 cm}
\subsection{Sensor set ($J$) configuration}
\label{subsec:Sensors_set}
For the system in \eqref{eq:dynamicSysExample}-\eqref{eq:matrices_dynamic}, we only design 9 sets of disturbed sensors (equal to the number of observers) based on our method. In more detail, according to Section~\ref{subsec:Multi-observer}, and \eqref{eq:dynamicSysExample}-\eqref{eq:matrices_dynamic}, the number of all possible subsets of disturbed sensors for $q=9$ (number of sensors) is
\begin{equation}
    \sum_{i=1}^q \left( \begin{matrix} q \\ i \end{matrix} \right) = 511.
    \label{eq:subsets_all}
\end{equation}
However, we are only interested in the "independent"  subsets that have detectable pair ($A, C_{J_j}$) according to Section~\ref{sec:methodology} which are 9 subsets as $J_{1} = \{ 1,2 \}$, $J_{2} = \{ 1,7 \}$, $J_{3} = \{ 1,9 \}$, $J_{4} = \{ 2,6 \}$, $J_{5} = \{ 2,8 \}$, $J_{6} = \{ 6,7 \}$, $J_{7} = \{ 6,9 \}$, $J_{8} = \{ 7,8 \}$, and $J_{9} = \{ 8,9 \}$. Thus $j \in \{1,\cdots, N \}$, $N = 9$ from the 511 subsets in \eqref{eq:subsets_all}, and the system can guarantee security up to two sensors under attack according to \eqref{eq:num_tolerable_attack}.

\vspace{-0.3 cm}
\subsection{Motion comfort assessment}
\label{subsec:Human_model}

ISO-2631:1997 \cite{an1997mechanical} provides guidelines for objectively measuring and evaluating human exposure to whole-body mechanical vibration and repeated shock. 
The guidelines suggest the consideration of two metrics: (1) Ride Comfort (RC) emphasizing the higher frequencies (mainly above 1 Hz); (2) MS emphasizing the lower frequencies (mainly below 1~Hz). The first metric (RC) reflects the immediate perception of motion discomfort through perceivable accelerations which are normally road induced but in this paper will result from forwards/rearward accelerations induced by sensor attacks and by acceleration and deceleration of the lead vehicle. The second metric (MS) reflects low frequency components which will induce motion sickness in particular in passengers and AV users taking the eyes off the road over longer periods.

The metrics are evaluated by combining the Root Mean Square (RMS) values of weighted accelerations ($a_{W_{o}}$), translational and rotational, measured at the vehicle's center of gravity.
More specifically, the RMS value of each acceleration is calculated as follows: 

\begin{equation}
\label{eq:accel}
	\textrm{RC}_{o}=  \bigg( \frac{1}{t}  \int_{0}^{t} a_{W_{o}}^2(s) ds \bigg)^{\frac{1}{2}},
\end{equation}

\noindent where $o$ is the acceleration type, either translational or rotational, while $a_{W_{o}}$ stands for the frequency weighted accelerations in the time domain.
The overall RC is calculated as: 


\begin{equation}
\textrm{RC} = \bigg( \sum k_{o}^2 \textrm{RC}_{o}^2 \bigg)^{\frac{1}{2}},
\label{eq:rc_index}
\end{equation}

\noindent where $k_{o}$ is a weighting factor for each term which can be found in the literature \cite{an1997mechanical}. 
This work addresses car following and hence only longitudinal accelerations ($o=X$) are considered both for comfort and motion sickness \cite{papaioannou2023impact}. 
For MS, the weighting filter for longitudinal $WP_{f_x}$ is used \cite{papaioannou2023impact},  which is similar to ISO and Motion Sickness Dose Value (MSDV).

\vspace{-0.3 cm}
\subsection{Assumed experimental parameters}
\label{sec:experimental_parameters}
Consider two homogeneous vehicles in a platoon i.e., $l = 2$ in Fig.~\ref{fig:ExampleModel}. We obtain the discrete-time system \eqref{eq:dynamicSysExample_dis} with sampling interval $T_s = 0.1$ seconds. The discrete-time system corresponds to the exact discretization of the continuous-time LTI system in \eqref{eq:dynamicSysExample}. Based on \cite{ploeg2013lp} we select the following CACC parameters and conditions. We select $h = 0.5$ seconds representing a relatively short following distance which can be string stable with CACC \cite{ploeg2013lp}. We select $\tau = 0.1$, $k_p = 0.2$, $k_d = 0.7$, $k_{dd} = 0.5$, $a_{\beta} = 1000$, $s_2 = 1$ m in \eqref{eq:dri}, and $K_r = 2$ and $C_r = 3$ (see Section~\ref{subsec:Reference_model_parameters_design}). We assumed that the system noise $w$ is i.i.d. uniformly distributed random vector normalized to satisfy noise bound $\mathcal{B}_{w} = 0$.
We simulate CACC where at each interval $k$, the desired acceleration of vehicle 1, i.e., $u_1(k)$ is transmitted from vehicle 1 to vehicle 2 via the communication network. Let $u_1(k) = 2 e^{-0.01 k}$, the initial relative velocity between vehicles 1 and 2 be 0.5 m/s, the initial spacing error be 0.1 m, and the initial velocity of vehicle 2 be 30 m/s. Therefore, $x_2(0) = [0.1,\;30,\;0,\;0,\;0.5,\;0]^{\top}$. The initial condition for each observer set of $\hat{x}_{J_j}(0)$, $j \in \{ 1, \cdots, N \}$, $N = 9$ (see Section \ref{subsec:Sensors_set}) is randomly chosen. Furthermore, we considered the initial condition of $\eta_{J_j}(0) = 0$, and $\beta_{J_j}(0) = 0.5$, $j \in \{ 1, \cdots, N \}$. 

\vspace{-0.3 cm}
\subsection{Benchmark method}
\label{subsec:Comparing_method}
We compare the performance of our work with that of Zhao et al.\cite{zhao2021resilient}, Ko~\cite{ko2021approach}, and Karmakar~\cite{karmakar2021assessing}. 

In \cite{zhao2021resilient}, the authors presented a recovery mechanism to confine the duration and frequency of adverse effects caused by attacks on platoon control. Meanwhile, a resilient platoon control protocol was proposed to achieve the internal stability of the Vehicular Cyber Physical Systems (VCPSs) under attacks. 
Although their methods successfully secure the system against attacks, as with most cybersecurity schemes, the dynamic systems experience significant excitation when the attacks begin, resulting in reduced comfort. For comparison, we use the same scheme as that used in Section V.B. Example 2 in~\cite{zhao2021resilient}.


In~\cite{ko2021approach}, an LSTM-based detection framework is proposed to identify malicious information attacks within connected adaptive cruise control (CACC) systems. The model captures temporal dependencies in vehicular data to detect anomalies effectively. Meanwhile, Karmakar et al.~\cite{karmakar2021assessing} present a deep learning-based approach that evaluates the trustworthiness of autonomous vehicles operating under cyberattack scenarios. While both methods demonstrate competent attack detection and mitigation capabilities, their reliability under diverse operating conditions remains insufficiently validated, which may hinder robust deployment in safety-critical environments.

\vspace{-0.1 cm}
\section{Simulation Results}

\vspace{-0.1 cm}
\subsection{Observer gain ($L_{J_j}$) design using LMI}
\label{subsec:LMI_L_K}

Having $N=9$, we solved all the $2N+1 = 19$ inequalities in \eqref{eq:lmi_A_P}-\eqref{eq:lmi_P} to find $P$, and $Z_{J_j}$, $j \in \{1, \dots, 9 \}$, using \texttt{gevp}
(Generalized eigenvalue minimization under LMI  constraints) function in MATLAB~\cite{nesterov1994interior}, which leads us to define all the observer gains $L_{J_j}$, $j \in \{1, \dots, 9 \}$ according to \eqref{eq:L_Jj_Design}.

\vspace{-0.3 cm}
\subsection{Reference model parameters ($K_r$, $C_r$) design}
\label{subsec:Reference_model_parameters_design}

According to $A_r^c$ in \eqref{eq:matrices_dynamic_dis}, and assuming that the eigenvalues of the reference model system are $\lambda_{r,1}$ and $\lambda_{r,2}$, the parameters of $A_r^c$ should be 
\begin{align}
    C_r &= -(\lambda_{r,1} + \lambda_{r,2}), \nonumber \\
    K_r &= \lambda_{r,1} \; \lambda_{r,2}.
    \label{eq:Reference_model_parameters}
\end{align}
The system is asymptotically stable ($A_r^c$ is Hurwitz) if $\lambda_{r,1} < 0$ and $\lambda_{r,2} < 0$. For our simulation, we consider $\lambda_{r,1} = -1$, and $\lambda_{r,1} = -2$ which results in $K_r = 2$ and $C_r = 3$. Note that the eigenvalues are selected customary while they are holding distinct negative real values \cite{khalil2002nonlinear}.

\vspace{-0.3 cm}
\subsection{Simulation scenario}
\label{subsec:Experiment_scenario}

Here, we study bounded attacks with three attack severities (magnitudes), ordered from high to low as 1) Critical, 2) Very Uncomfortable, and 3) Uncomfortable. 
The Critical Safety attacks are designed as intense attacks that result in collisions when no countermeasures are taken, the Very Uncomfortable attacks are designed such that they lead to considerable discomfort, and the Uncomfortable attacks' design achieves considerably more power than common sensor noise while it does not result in collisions or major discomfort. To generate Critical Safety Attacks, the attack amplitude was increased to ensure that at least one collision occurred in each attack. Very Uncomfortable Attacks were then derived by gradually reducing the attack amplitude from the Critical Safety Attack level until no collisions were observed. Finally, Uncomfortable Attacks were formulated by adjusting the attack amplitude such that the overall MSDV$_x$ and RC values fell within the predefined uncomfortable range \cite{ISO2631}. 
\begin{figure}[t]
	\centering
	\includegraphics[width= 1\linewidth]{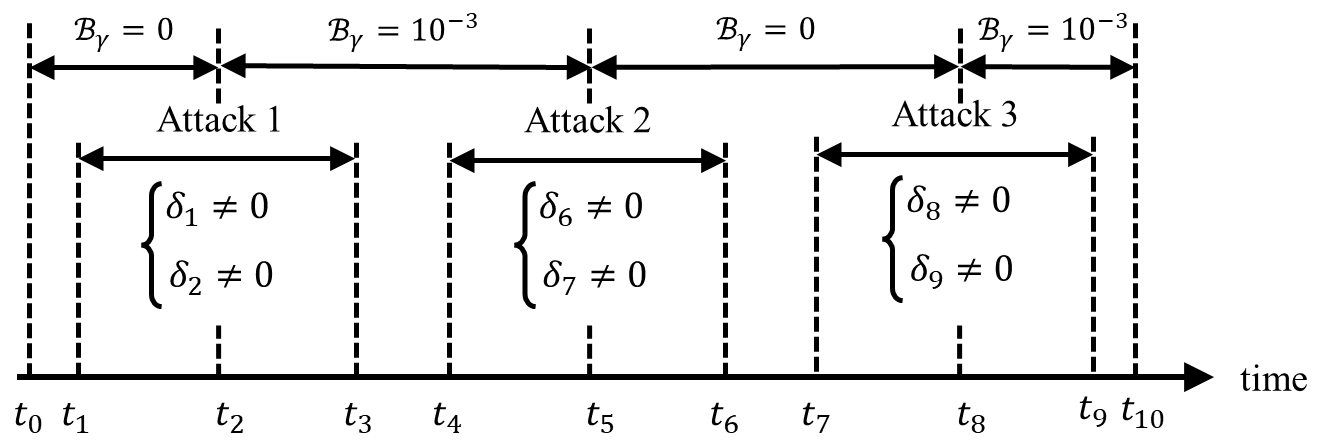}
	\caption{Designed simulation scenario.}
	\label{fig:Timing_Scenario}
    \vspace{\VDis}
\end{figure}

Within each attack magnitude case, three different attacks are simulated. In each of the attacks, two of the sensors are hacked at the same time where the hacked sensors are different from one attack to the other attack. According to Fig.~\ref{fig:Timing_Scenario}, for two parts of the simulation (from $t_2$ to $t_5$ and from $t_8$ to $t_{10}$), normal sensor noise $\gamma_i$ is nonzero as a realistic condition and is i.i.d. uniformly distributed random vectors normalized to satisfy noise bounds $\mathcal{B}_{\gamma} = 10^{-3}$. For the rest of the simulation, we set $\mathcal{B}_{\gamma} = 0$. In addition, in Attack 1, sensors $y_1$ and $y_2$ are under FDI attack between $t_1$ and $t_3$. Similarly, 
in Attack 2 which is between $t_4$ and $t_6$, sensors $y_6$ and $y_7$ are under FDI attack. Furthermore, in Attack 3, sensors $y_8$ and $y_9$ are under FDI attack between $t_7$ and $t_9$. 

Both Attack 1 and Attack 2 are white noise, and Attack 3 is a repeatedly on-off switching white noise where the RMS value of each of attack per each scenario is reported in TABLE~\ref{tab:RMS_attack_scenario}. Attack 3 is repeatedly (de)activated by an activation logic as follows.
\begin{table}[tb]
	\caption{RMS value of designed attacks perseverity level.}
	\label{tab:RMS_attack_scenario}
	\centering
	\resizebox{\linewidth}{!}{

\input{RMS_attack_scenario}}
    \vspace{\VDis}
\end{table}
\begin{align}
    \begin{cases}
        \delta_8(k) = 0, \delta_9(k) = 0, \;\; \textrm{if} \; &\{\textrm{floor}(kT_s) = 2l \;|\; l\in \mathbb{Z} \},  \\
        \delta_8(k) \neq 0, \delta_9(k) \neq 0 &\textrm{otherwise}.
    \end{cases}
    \label{eq:attack3}
\end{align}

By setting $\mathcal{B}_{\gamma} = 0$ between $t_5$ and $t_8$, we investigate the asymptotical stability of the framework in the presence of attack and absence of normal sensor noise. 
We performed a 30 minutes simulation where $t_0 = 0$s, $t_1 = 60$s, $t_2 = 300$s, $t_3 = 540$s, $t_4 = 660$s, $t_5 = 900$s, $t_6 = 1140$s, $t_7 = 1260$s, $t_8 = 1500$s, $t_9 = 1740$s, and $t_{10} = 1800$s.

\vspace{-0.3 cm}
\subsection{Results: Steady state platooning}
\label{subsec:results}
To evaluate our method comprehensively, we evaluate the scenario of steady state platooning with the three sequential attacks defined in Section~\ref{subsec:Experiment_scenario}. We compare four conditions as follows from securing safety and securing comfort points of view:
\begin{itemize}
    \item \textbf{Condition~1 (Insecure controller~\cite{ploeg2013lp} without attack}): Depicts the performance of the controller in \eqref{eq:cont111} in the absence of attack ($\delta_i = 0, i\in \{ 1, \dots, 9\}$). Note that, here, for the variables measured with multiple sensors we use the average of the measured values for a variable to design the control signal in \eqref{eq:cont111}.
    \item \textbf{Condition~2 (Insecure controller~\cite{ploeg2013lp} with attack}): Depicts the performance of the controller in \eqref{eq:cont111} but not using any cyber attack security approach along with the three designed attacks in Section~\ref{subsec:Experiment_scenario}. Same as Condition 1, for the variables measured with multiple sensors we use the average of the measured values for a variable to design the control signal in \eqref{eq:cont111}.
    \item \textbf{Condition~3 (Zhao's method~\cite{zhao2021resilient} with attack}): Depicts the performance of the security method in \cite{zhao2021resilient} (explained in Section~\ref{subsec:Comparing_method}) along with the designed attacks in Section~\ref{subsec:Experiment_scenario}.
    \item \textbf{Condition~4 (Our method with attack}): Depicts the performance of our security method explained in Section~\ref{sec:methodology} using controller \eqref{eq:dynamicSysExample_dis}, along with the designed attacks in Section~\ref{subsec:Experiment_scenario}. The controller in \eqref{eq:dynamicSysExample_dis} is the secure version of the controller in \eqref{eq:cont111} using our method to reject attacked sensors.
\end{itemize}

\subsubsection{Securing safety} 
\begin{figure}[tb]
		\centering
    \subfloat{\includegraphics[width = 1\linewidth]{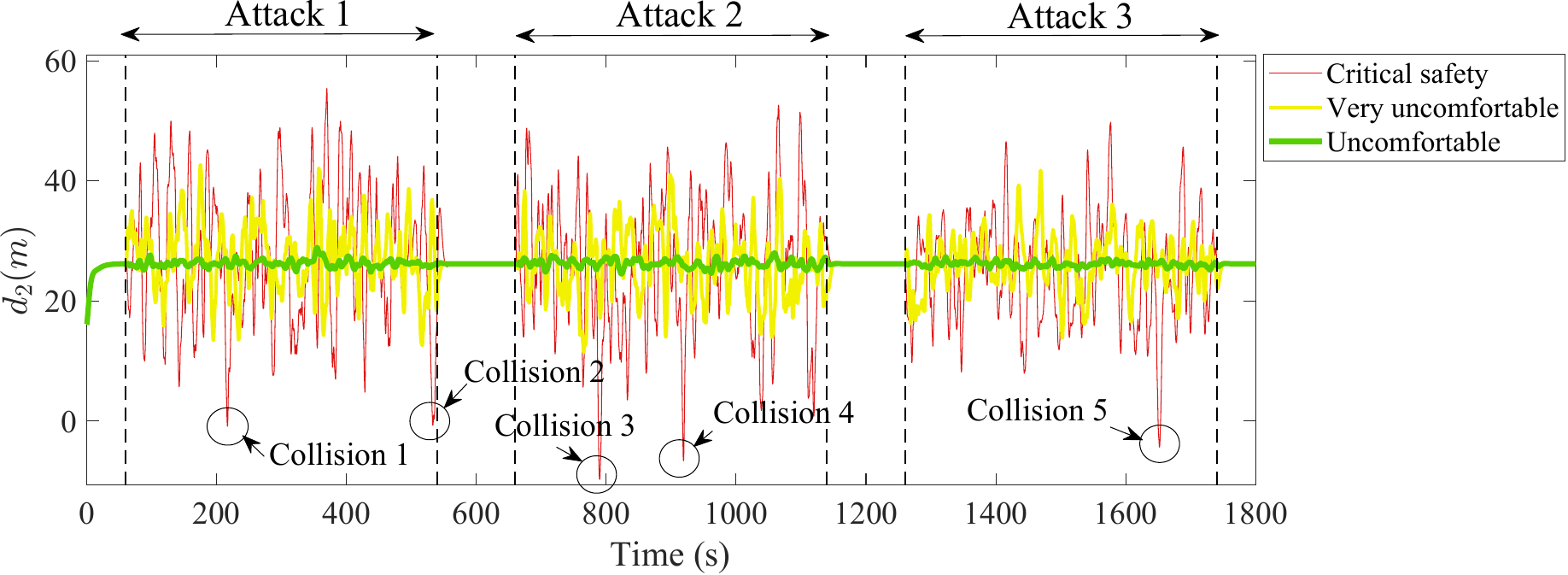}}\hspace{0 cm} 
	\caption{Following distance ($d_2$) between the Vehicle $2$ and its preceding Vehicle $1$ per each attack severity level by the insecure platooning controller in \eqref{eq:cont111} (Condition 2).}
	\label{fig:results_distance}
    \vspace{\VDis}
\end{figure}

Fig.~\ref{fig:results_distance} shows the followed distance ($d_2$) in Condition 2 (Insecure controller~\cite{ploeg2013lp} with attack) where $d_2<0$ defines the collisions. The designed Critical Safety attacks result in five collisions, the Very Uncomfortable attacks lead to 30 m oscillations of the $d_2$ and no collision, and the Uncomfortable attacks lead to 5 m oscillations of the $d_2$ and no collision. TABLE~\ref{tab:Safety_SteadyState} summarizes the main results, namely the Number of Collisions (NC) and RMS of the spacing error (RMS$_{e_2}$) per severity level. Apparently both {Zhao's method~\cite{zhao2021resilient}} (Condition 3) and our method (Condition 4) avoid all collisions but our method achieves a much lower spacing error which is almost as low as in the case without attack. 

\begin{figure}[tb]
		\centering
    \subfloat[]{\includegraphics[width = 1\linewidth]{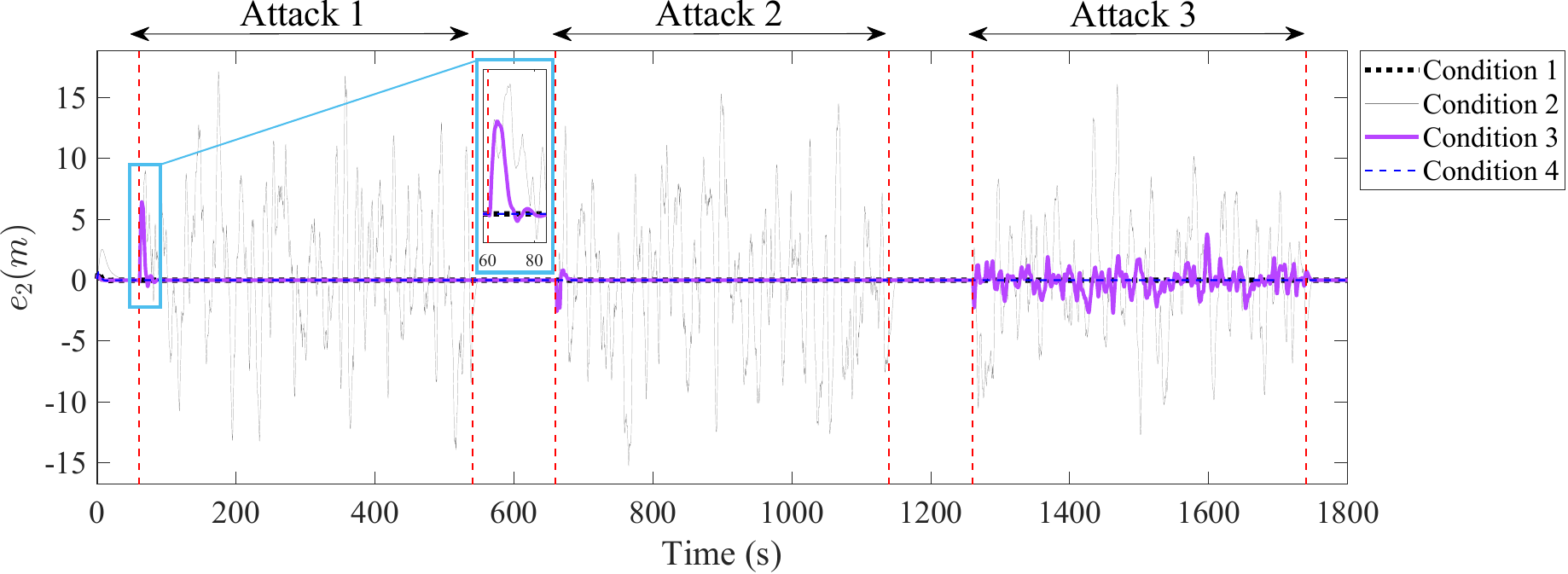}}\hspace{0 cm} 
    \subfloat[]{\includegraphics[width =  1\linewidth]{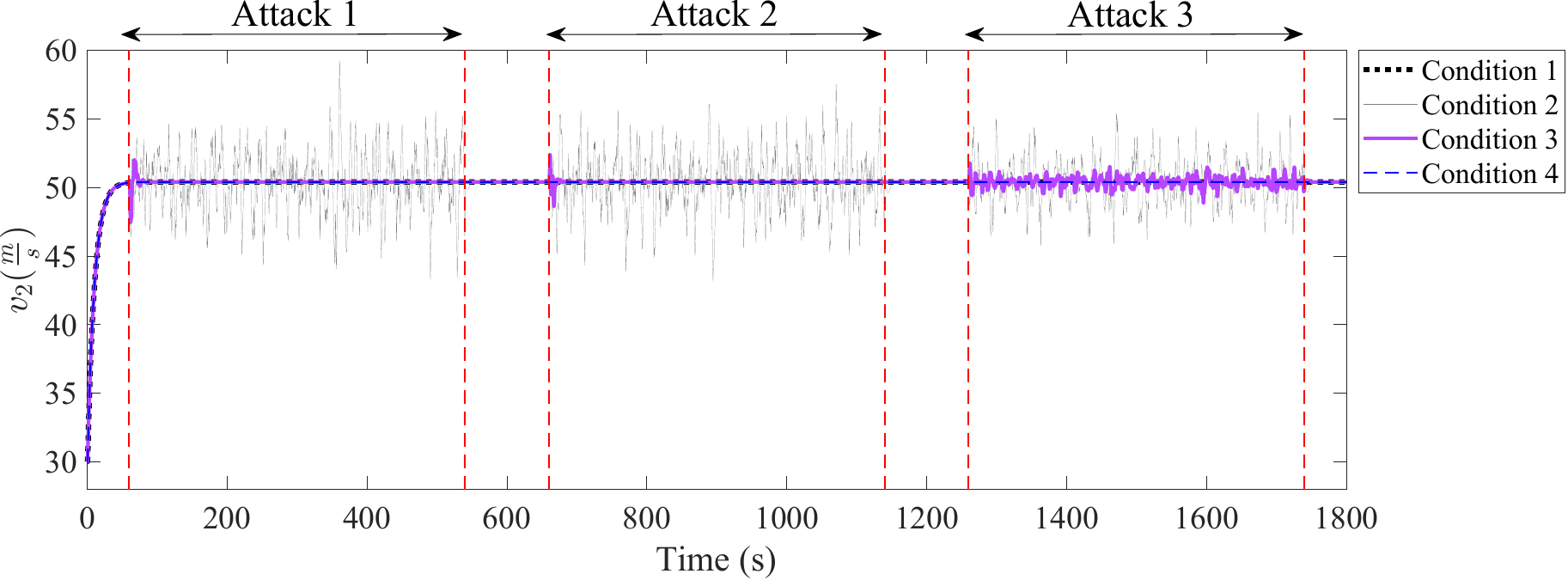}}\hspace{0 cm}
    \subfloat[]{\includegraphics[width =  1\linewidth]{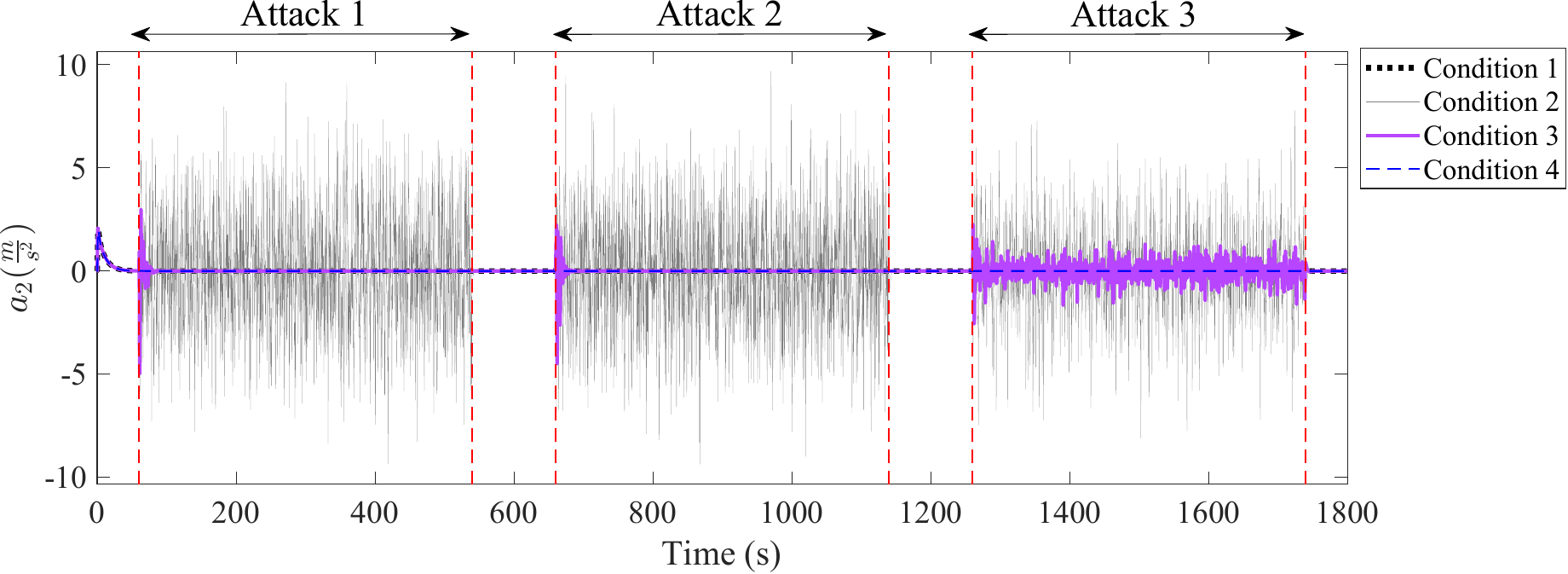}}\hspace{0 cm}
    \subfloat[]{\includegraphics[width =  1\linewidth]{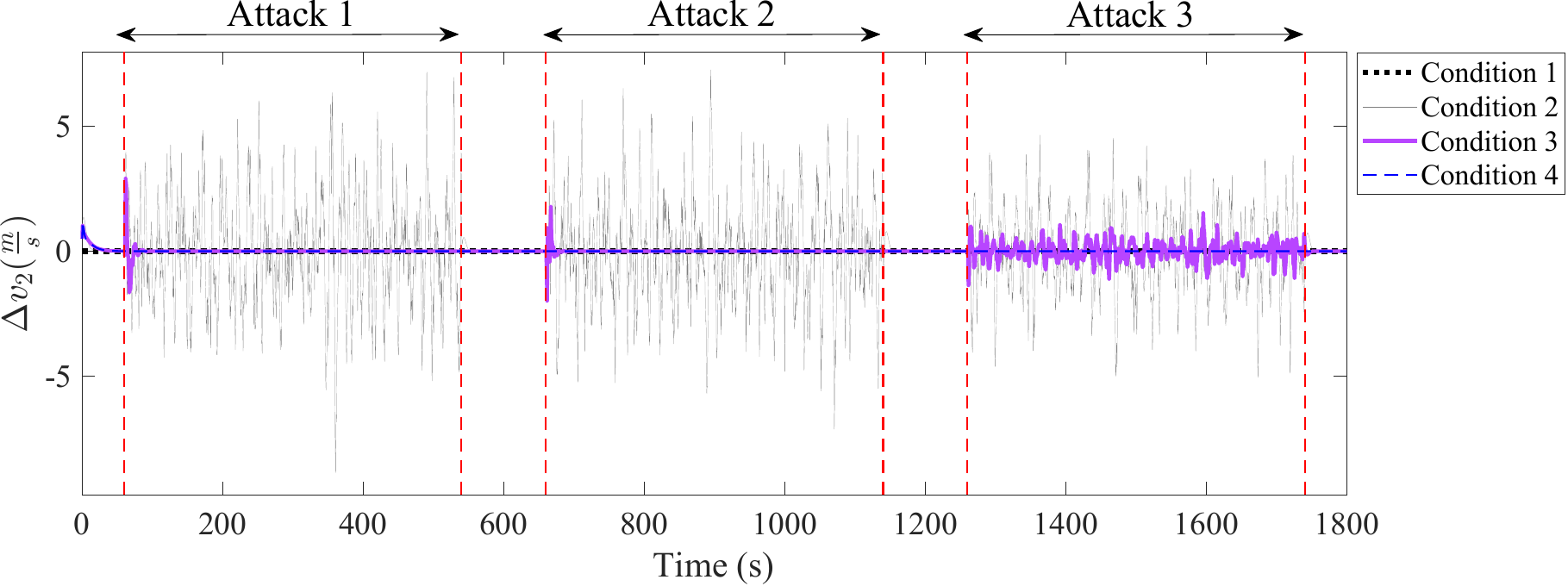}}\hspace{0 cm}   
	\caption{Control performance for the Very Uncomfortable attack severity: a) Spacing error ($e_2$), b) Velocity ($v_2$), c) Acceleration ($a_2$), d) Relative velocity ($\Delta v_2$).}
	\label{fig:results_states_Comparision}
    \vspace{\VDis}
\end{figure}

Fig.~\ref{fig:results_states_Comparision} illustrates the attack detection dynamics as function of time for the continuous Attacks 1 and 2 and the rapidly switching Attack 3. Obviously, the CACC without any security approach (Condition 2) has the highest vulnerability to sensor attacks compared to Conditions 3 and 4. 
In Fig.~\ref{fig:results_states_Comparision}, Zhao's method~\cite{zhao2021resilient} (Condition 3) presents asymptotic stability of the estimation error against Attack 1 and Attack 2, while at the beginning of the attacks, namely at $t_1$ for Attack 1 and $t_4$ for Attack 2, the system states are excited for a while. This is more evident in the zoomed-in area at time $t_1$ in Fig.~\ref{fig:results_states_Comparision}(a), where it takes approximately 20 seconds for Condition 3 to reject the attack impact on the system states. In Attack 3, Zhao's method~\cite{zhao2021resilient} (Condition 3) fails to secure the system, and Attack 3 successfully disturbs the system states due to the states' excitations in every two seconds defined in \eqref{eq:attack3}. The failure can be addressed more clearly in TABLE~\ref{tab:Safety_SteadyState} where the resulting RMS$_{e_2}$ = 0.58 (m), which defines a considerable error safe distance ($d_{r,2}$).  
However, based on Fig.~\ref{fig:results_states_Comparision}, our method (Condition 4) could successfully secure the safety of vehicle platooning control by almost completely  rejecting the impact of all three attacks. For instance, Condition 4 rejects the attack impact almost immediately, in contrast to Zhao's method~\cite{zhao2021resilient}, as illustrated in the zoomed-in area at time $t_1$ in Fig.~\ref{fig:results_states_Comparision}(a). In terms of safety, our method (Condition 4) resulted in RMS$_{e_2}$ = 0.03 (m) similar to the best performance, i.e. Condition 1 without any attacks, and about 94\% less than the RMS$_{e_2}$ of Zhao's method~\cite{zhao2021resilient} (Condition 3).

\begin{figure}[tb]
		\centering
    \subfloat[]{\includegraphics[width =  1\linewidth]{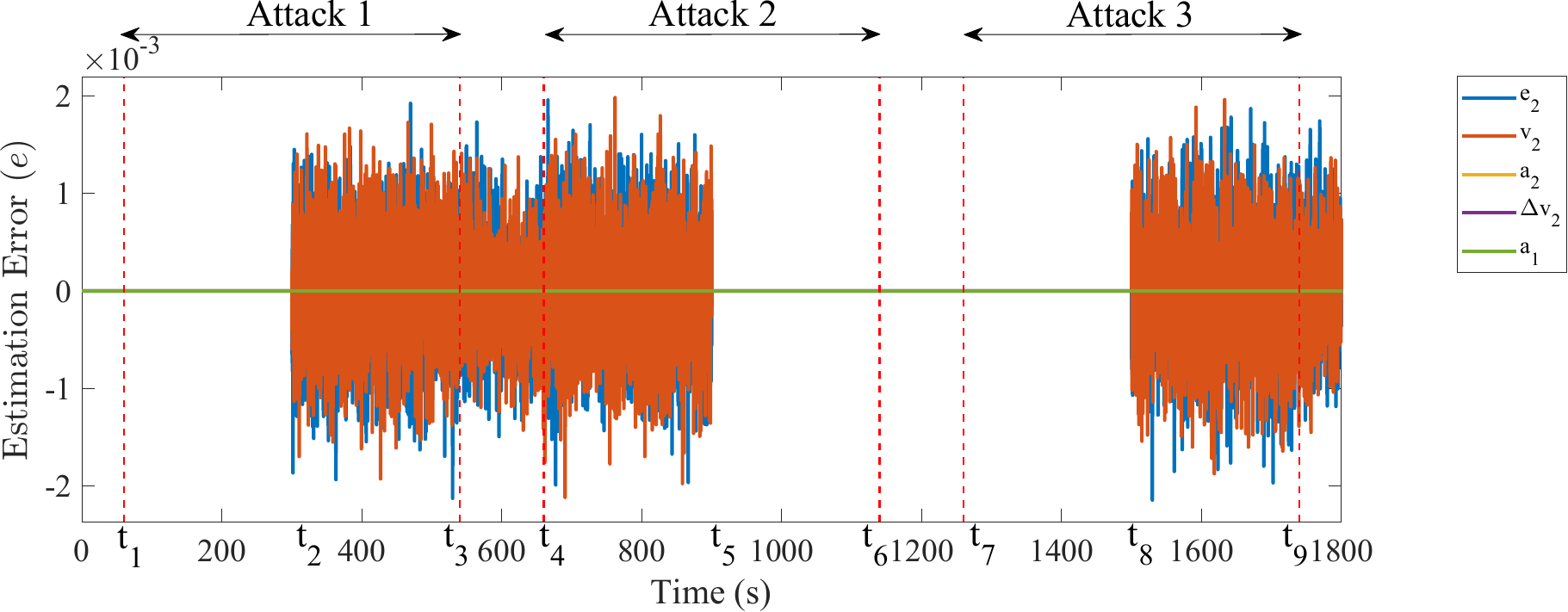}}\hspace{0 cm}
    \subfloat[]{\includegraphics[width =  1\linewidth]{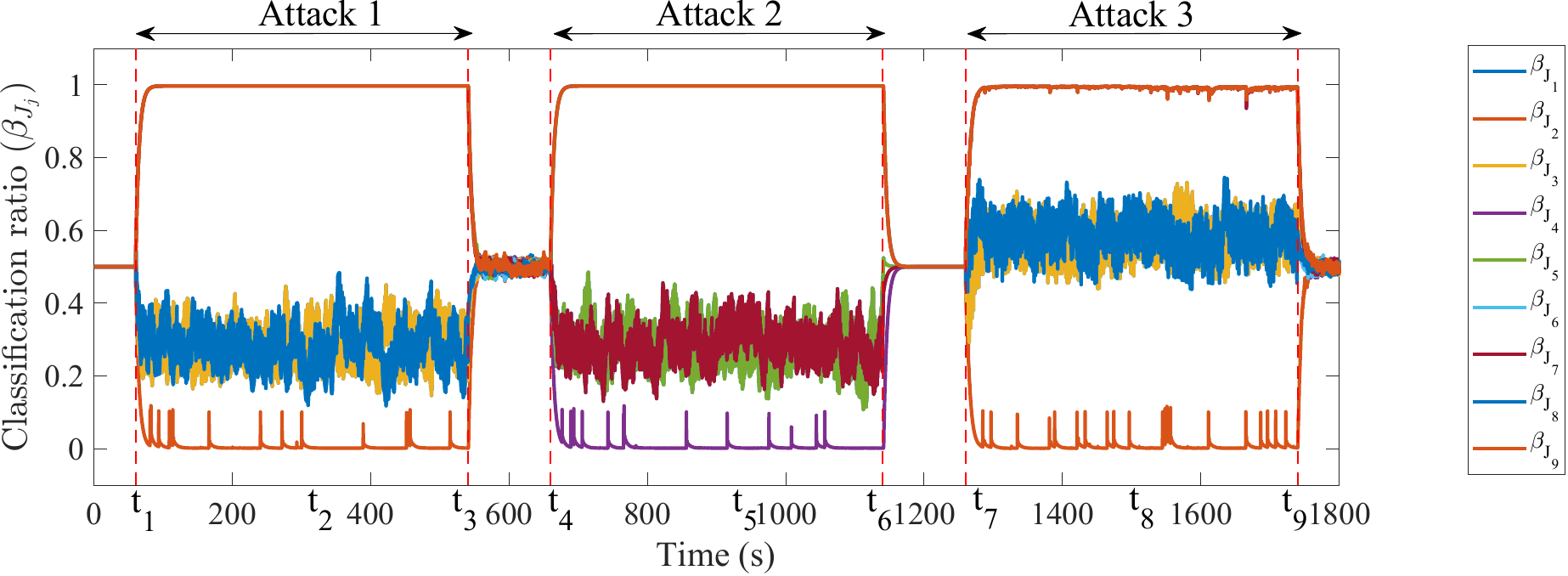}}\hspace{0 cm}
    \subfloat[]{\includegraphics[width =  1\linewidth]{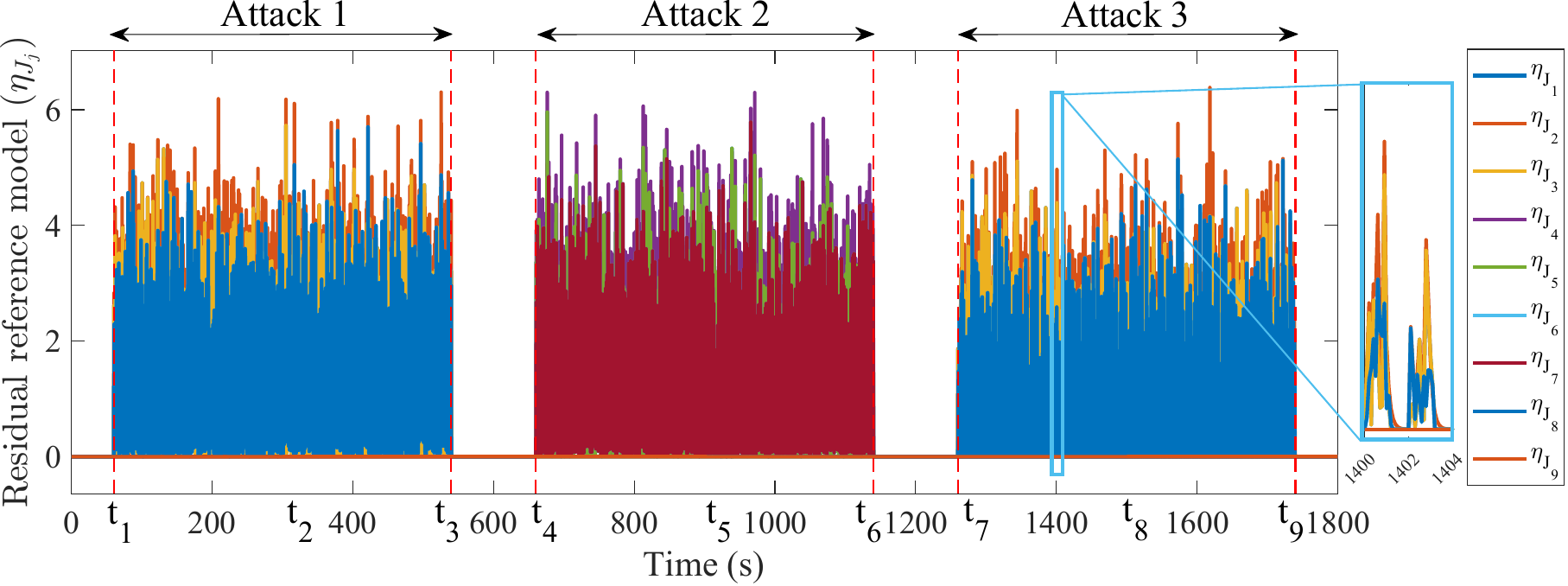}}\hspace{0 cm}
	\caption{The optimal-coupling-observer framework's components' performance (Condition 4): a) Estimation error, b) classification ratio, c) residual reference model.}
	\label{fig:results_variables_framework}
    \vspace{\VDis}
\end{figure}

The asymptotic stability with no states' excitation can be investigated more precisely in Fig.~\ref{fig:results_variables_framework}(a), where the estimation error in \eqref{eq:sum_error} is depicted. Between $t_2$ and $t_5$ and between $t_8$ and $t_{10}$, the error is only affected by the normal sensor noise which is inevitable and cannot be rejected by any existing methods due to the best of our knowledge. Furthermore, between $t_0$ and $t_2$ and between $t_5$ and $t_8$, the estimation error in Condition 4 (contrary to Condition 3) not only rejects the impact of attacks with no states' excitation but also is asymptotically stable and converges to zero which reflects the strong stability of the framework. 
Fig.~\ref{fig:results_variables_framework} (b) and (c) draw the performance of each framework component. In Fig.~\ref{fig:results_variables_framework}(b) the classification ratio's \eqref{eq:classification_ratio} performance to the noise and the attacks is depicted. The classification ratio could classify each observer based on the degree of affectedness to the attack. In each attack, the classification ratio includes three regions around 1, 0.6, 0.3, and 0. Apparently, the framework is using the observer with the greatest value for classification ratio (ideally equal to 1) to estimate the system states.

Fig.~\ref{fig:results_variables_framework}(c) shows that the residual reference model in \eqref{eq:matrices_dynamic_dis} is asymptotically stable in the usual sense in the absence of noise and attack, namely between $t_6$ and $t_7$, and also can be correctly excited in the presence of three of the attacks. In the rapidly switching Attack 3, which was a bottleneck for Zhao's method~\cite{zhao2021resilient} (Condition 3), it is clear that the residual reference model is correctly excited every two seconds, namely at $t = 1400s$, and $t = 1402s$ which are the start of the on-mode of the Attack 3 in~\eqref{eq:attack3}.

Regarding the Critical Safety and Uncomfortable severity levels in TABLE~\ref{tab:Safety_SteadyState}, the same conclusion as Very Uncomfortable severity level emerges. 
Condition 2 is the most vulnerable to attacks, Condition 3 is considerably affected by Attack 3, and Condition 4 (our method) performs similar to Condition 1. However, Condition 2 under Critical Safety attacks results in instability of the Vehicle 2 states with five collisions shown in Fig.~\ref{fig:results_distance} (while they are bounded during all of the Attacks) which strongly threats safety.

\subsubsection{Securing comfort}

\begin{figure}[tb]
		\centering
    \subfloat[]{\includegraphics[width = 0.49\linewidth]{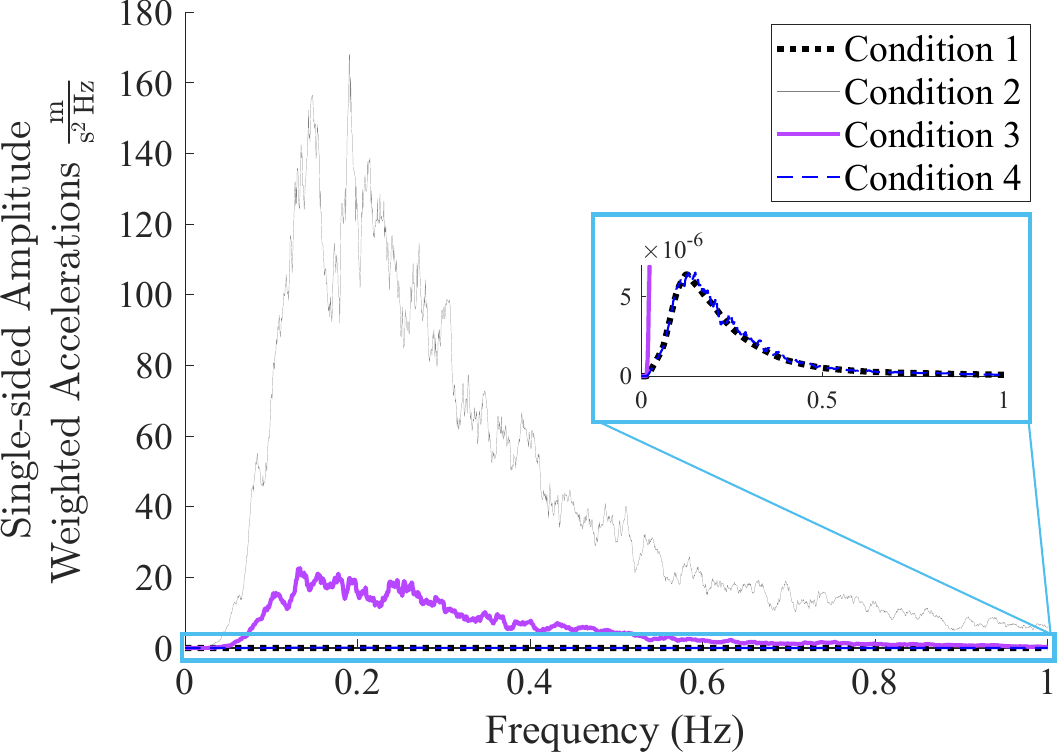}}\hspace{0 cm} 
     \subfloat[]{\includegraphics[width = 0.49\linewidth]{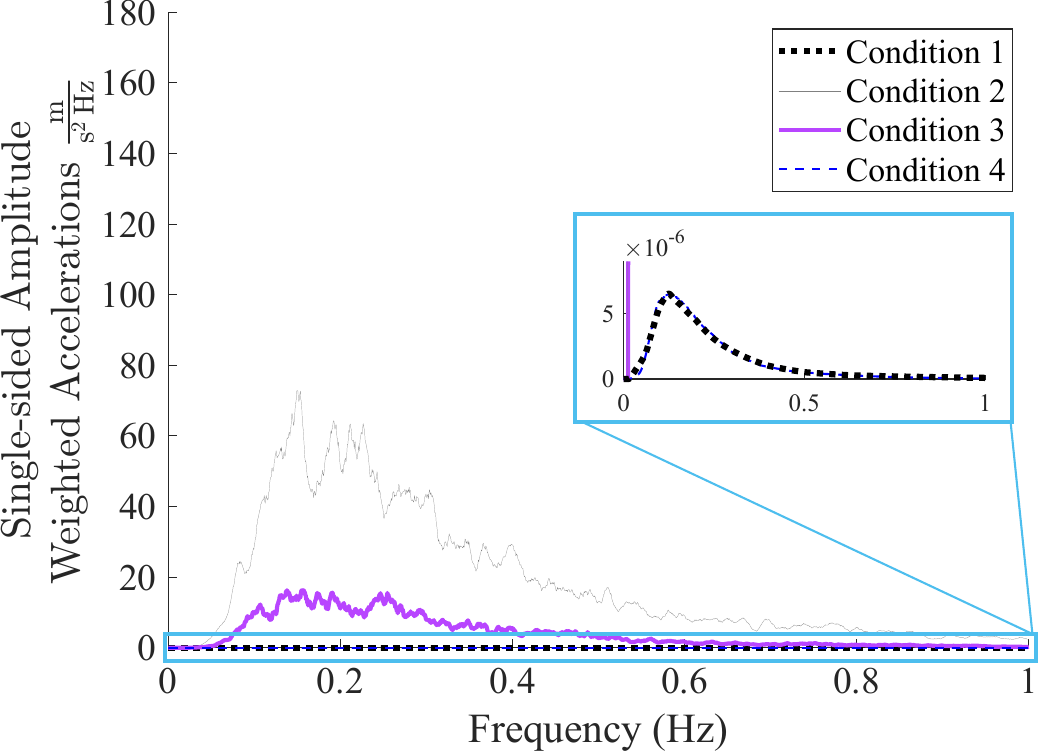}}\hspace{0 cm} 
    \caption{Frequency analysis for the Very Uncomfortable attack severity: (a) Vehicle 2, (b) Vehicle 10 in platooning.}
	\label{fig:FreqAnalysis}
    \vspace{\VDis}
\end{figure}

To assess the occupants' comfort, the longitudinal accelerations of Vehicle 2 in the platoon with Very Uncomfortable severity level are compared in the frequency domain in Fig.~\ref{fig:FreqAnalysis}(a). In addition, the objective comfort scores (MSDV$_x$ reflecting motion sickness and RC reflecting ride comfort, Section~\ref{subsec:Human_model}) based on the vehicle's longitudinal acceleration are reported in TABLE~\ref{tab:Comfort_SteadyState} per each severity level.

In condition 2 without security method the most severe "Critical Safety" attack with $\textrm{RC}=3.08$ $\frac{m}{s^2}$ is indeed classified as extremely uncomfortable ($\textrm{RC}>2$ $\frac{m}{s^2}$ according to ISO2361) whereas the "Very Uncomfortable" attack with $\textrm{RC}=1.59$ $\frac{m}{s^2}$ is indeed classified as very uncomfortable ($1.25<\textrm{RC}<2.5$ $\frac{m}{s^2}$ according to ISO2631).

Fig.~\ref{fig:FreqAnalysis} illustrates the longitudinal accelerations as filtered with regards to motion sickness (Section~\ref{subsec:Human_model}).
According to Fig.~\ref{fig:FreqAnalysis}(a), the highest of the longitudinal accelerations is identified in Condition 2. 
This is consistent across the whole frequency range, relevant for motion sickness i.e., 0 to 1 Hz. 
This shows that the system with insecure controller is sensitive to attacks in the frequency range which affects the occupants' sickness drastically (i.e., around 0.2 Hz).
Meanwhile, the lowest power is identified in Conditions 1 and 4. 
The performance of Condition 3 shows that although the compared method in Section~\ref{subsec:Comparing_method} could reduce the maximum power for about 87\% compared to the insecure controller in Condition 2, it is considerably greater than the almost zero power (0.01 ($\frac{m}{s^2 Hz}$)) in Condition 4 as well as no attack scenario in Condition 1. 
Furthermore, our method in Condition 4 has an almost identical response to no attack scenario in Condition 1, illustrating that the impact of the attack was almost completely zeroed out. 
As a result, no additional sickness would occur to the occupants. 

According to TABLE~\ref{tab:Comfort_SteadyState}  for Critical Safety severity level, the generated MSDV$_x$ = 10.41 and RC = 0.29 $\frac{m}{s^2}$ in Condition 3 are reduced about 89\% and 91\% compared to the insecure controller in Condition 2, respectively, while they are considerably higher than MSDV$_x$ = 0.0028 and RC = 4e-5 $\frac{m}{s^2}$ for Condition 4 as well as no attack scenario in Condition 1. In addition, for Very Uncomfortable severity level, the MSDV$_x$ = 6.57 and RC = 0.16 $\frac{m}{s^2}$ in Condition 3 decreased about 86\% and 89\% compared to Condition 2, respectively, however, they are higher than MSDV$_x$ = 0.0028 and RC = 4e-5 $\frac{m}{s^2}$ for Condition 4 as well as no attack scenario in Condition 1. Finally, for Uncomfortable severity level, the generated MSDV$_x$ = 1.01 and RC = 0.02 $\frac{m}{s^2}$ in Condition 3 dropped about 78\% and 84\% compared to the insecure controller in Condition 2, and increased drastically compared to Condition 1 with MSDV$_x$ = 0.0028 and RC = 4e-5 $\frac{m}{s^2}$, respectively. As a result, Zhao's method~\cite{zhao2021resilient} could generally reduce MS and RC at all severity levels while it could not completely secure the occupant's comfort. 
\begin{table}[tb]
	\caption{Objective assessment of safety for steady state maneuver (attack 1-3) based on Number of Collisions (NC) and RMS of the spacing error in meters (RMS$_{e_2}$).}
	\label{tab:Safety_SteadyState}
	\centering
	\resizebox{\linewidth}{!}{

\input{Safety_SteadyState}}
    \vspace{\VDis}
\end{table}
\begin{table}[tb]
	\caption{Objective assessment of comfort for steady state maneuver (attack 1-3) based on Motion Sickness (MSDV$_x$) and Ride Comfort (RC) indexes.}
	\label{tab:Comfort_SteadyState}
	\centering
	\resizebox{\linewidth}{!}{

\input{Comfort_SteadyState}}
    \vspace{\VDis}
\end{table}
On the contrary and for all severity levels, our method (Condition 4) illustrated almost zero MS metric by canceling out the impact of the attack providing similar comfort with no attack scenario in Condition 1. 
There was a small increase ($\sim$ 4\%) for MSDV$_x$ compared to Condition 1, which was negligible due to the insignificant amplitude of the metric. 

\vspace{-0.3 cm}
\subsection{Platooning with 10 vehicles}

To further evaluate the scalability and robustness of the proposed method, we simulated a platoon of 10 vehicles, where Vehicle 1 is the lead vehicle and Vehicle 10 is the tail vehicle. As above, an attack was launched on Vehicle 2, following the scenario illustrated in Fig.\ref{fig:Timing_Scenario}. Now 8 additional vehicles were simulated to demonstrate attack attenuation in the platoon. A comparison of the single-sided amplitude-weighted acceleration spectra for Vehicle 2 and Vehicle 10, shown in Fig.\ref{fig:FreqAnalysis}(a) and Fig.\ref{fig:FreqAnalysis}(b), respectively, reveals notable differences in system response across control strategies. Zhao’s method\cite{zhao2021resilient} (Condition 3) reduced the peak amplitude by approximately 22\%, while the insecure controller in Condition 2 achieved a more substantial reduction of about 56\%. Remarkably, our proposed method (Condition 4) maintained performance nearly identical to the no-attack baseline (Condition 1), demonstrating high resilience to the injected disturbance.

These results highlight the string stability of the controller defined in \eqref{eq:dynamicSysExample_dis} used for Condition 1,2, and 4, which effectively suppresses disturbance propagation along the platoon. In contrast, Zhao’s method exhibited limitations under the repeatedly on-off switching attack scenario, where spacing errors were neither amplified nor sufficiently damped, thereby affecting downstream vehicle performance. Our method preserved the ride comfort of the tail vehicle at a level comparable to that of the second vehicle, underscoring its effectiveness in maintaining platoon stability and passenger comfort under adversarial conditions.

\vspace{-0.3 cm}
\subsection{State-of-Art reliability comparison}
To further compare the reliability of Zhao's method~\cite{zhao2021resilient} (Condition~3) with our method (Condition~4), we study variation of two metrics of False positive (FP) and F1-Score by running the scenario in Section~\ref{subsec:results} for 100 times and with different attack amplitudes from 0.0001 to 300. FP defines the number of intervals that the method mistakenly did not use the most reliable
observer, and F1-Score estimates the quality of resiliency under attacks. 


Since our method does not explicitly classify an observer as reliable or compromised but instead selects the most reliable observer at each time step, a FP occurs when the method selects a compromised observer. In such cases, the other, more reliable observers are incorrectly ignored, leading to a False Negative (FN). Consequently, the number of time intervals with FP is equal to those with FN. By the same reasoning, the number of time intervals for True Negative (TN) and True Positive (TP) are also equal. Since TP = TN and FP = FN, the evaluation metrics—F1-score, Precision, and Recall—yield the same value.
\begin{figure}[tb]
		\centering
    \subfloat{\includegraphics[width =  1\linewidth]{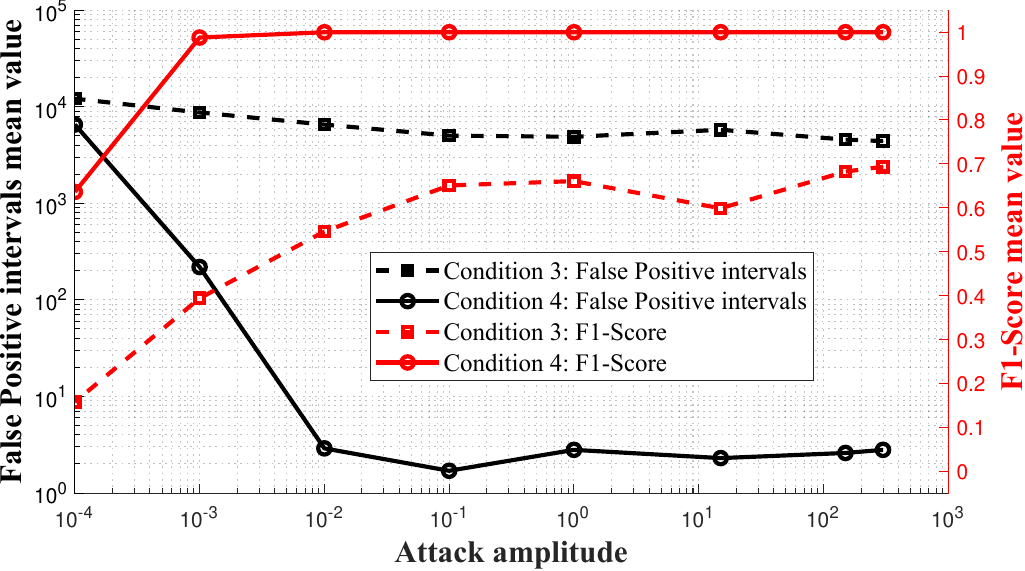}} 
	\caption{Reliability of attack detection mechanisms. 
    }
	\label{fig:reliability_of_attack}
    \vspace{\VDis}
\end{figure}

The reliability measures as function of attack amplitude  are depicted in Fig.~\ref{fig:reliability_of_attack}. From Fig.~\ref{fig:reliability_of_attack}(left axis) both methods have high FP mean values for attack amplitudes below or equal to 0.001. The reason is that the attack amplitude is below the noise amplitude and thus both methods are practically disabled. For attack amplitudes above 0.001, the FP mean value for our method (Condition 4) is around 2, while this is around 7000 for Zhao's method~\cite{zhao2021resilient} (Condition~3). This is because Zhao's method~\cite{zhao2021resilient} (Condition~3) fails to secure the system in presence of the on-off Attack 3 as discussed in Section~\ref{subsec:results}. In practice, the FP around 2 means that attacks of relevant magnitude are recognised in around two time steps after the start of the attack (0.2 seconds) and that the end of the attack is detected similarly, while during the attack the detection remains correct. This shows that the attack detection with our method is considerably quicker than Zhao's method~\cite{zhao2021resilient}. Fig.~\ref{fig:reliability_of_attack}(right axis) gives the same conclusion for the attacks above noise amplitude (0.001), where our method (Condition 4) with mean F1-Score of 0.999 is about 53\% more reliable than Zhao's method~\cite{zhao2021resilient} (Condition~3) with mean F1-Score of 0.65. The main reason for the reliability drop by Zhao's method~\cite{zhao2021resilient} (Condition~3) results from the vulnerability of the method to on-off Attack 3.
\begin{remark}
    For a broader comparison with SOTA cybersecurity frameworks, we evaluate the performance of our method (Condition 4) alongside Zhao's method~\cite{zhao2021resilient} (Condition 3), as well as the approaches by Ko~\cite{ko2021approach} and Karmakar~\cite{karmakar2021assessing}, in terms of F1-score for Attack 1 and Attack 2. Attack 3 is not addressed by the methods of Ko~\cite{ko2021approach} and Karmakar~\cite{karmakar2021assessing} and therefore excluded for this comparison.
Considering only Attack 1 and 2, our method (Condition 4) achieves an F1-score of 0.999, outperforming Condition 3 (F1-score = 0.97), as well as the methods by Ko~\cite{ko2021approach} and Karmakar~\cite{karmakar2021assessing} (both with F1-score = 0.96).
\end{remark}

\vspace{-0.3 cm}
\subsection{Results: Braking maneuver}
\label{subsec:results_Braking}
\begin{figure}[tb]
		\centering
    \subfloat[]{\includegraphics[width = 1\linewidth]{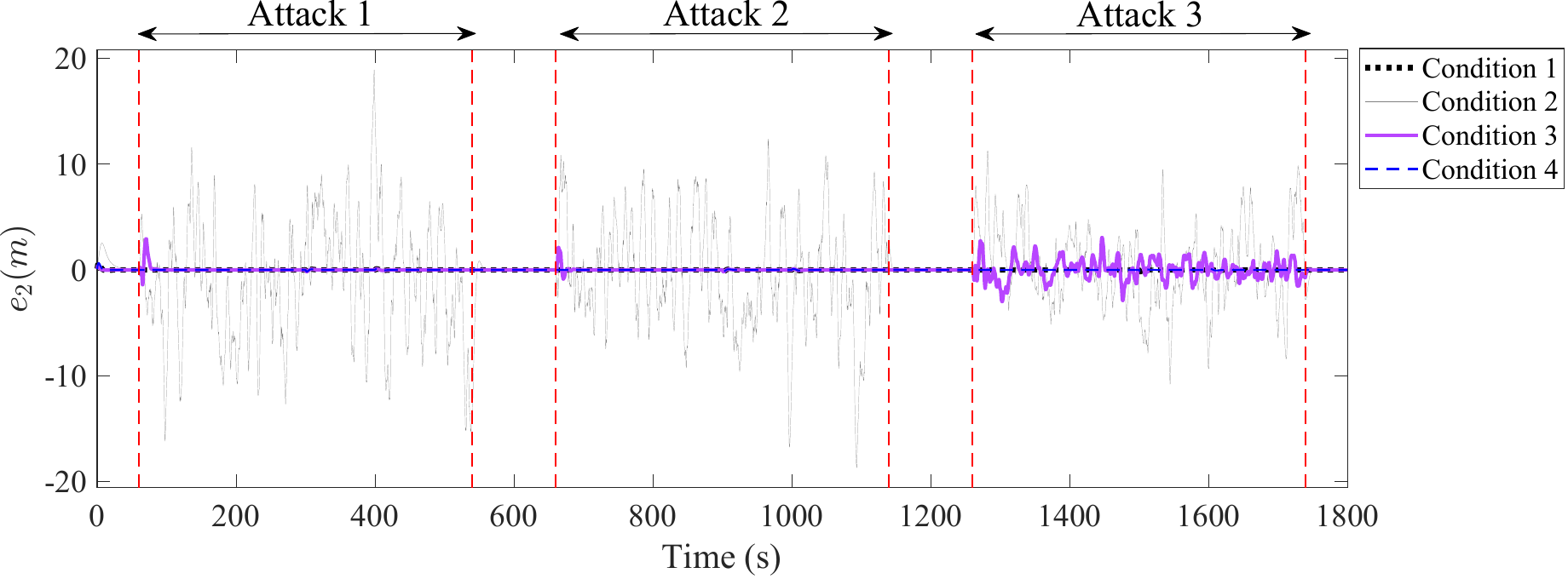}}\hspace{0 cm} 
    \subfloat[]{\includegraphics[width =  1\linewidth]{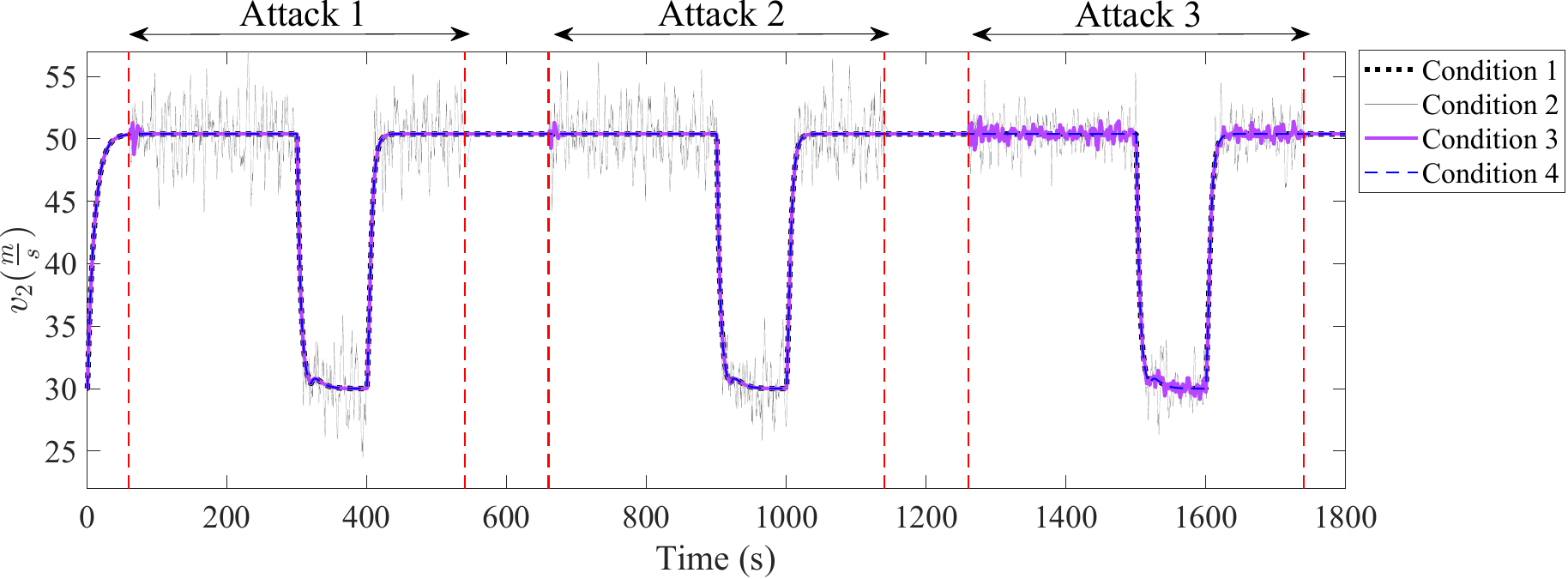}}\hspace{0 cm}
    \subfloat[]{\includegraphics[width =  1\linewidth]{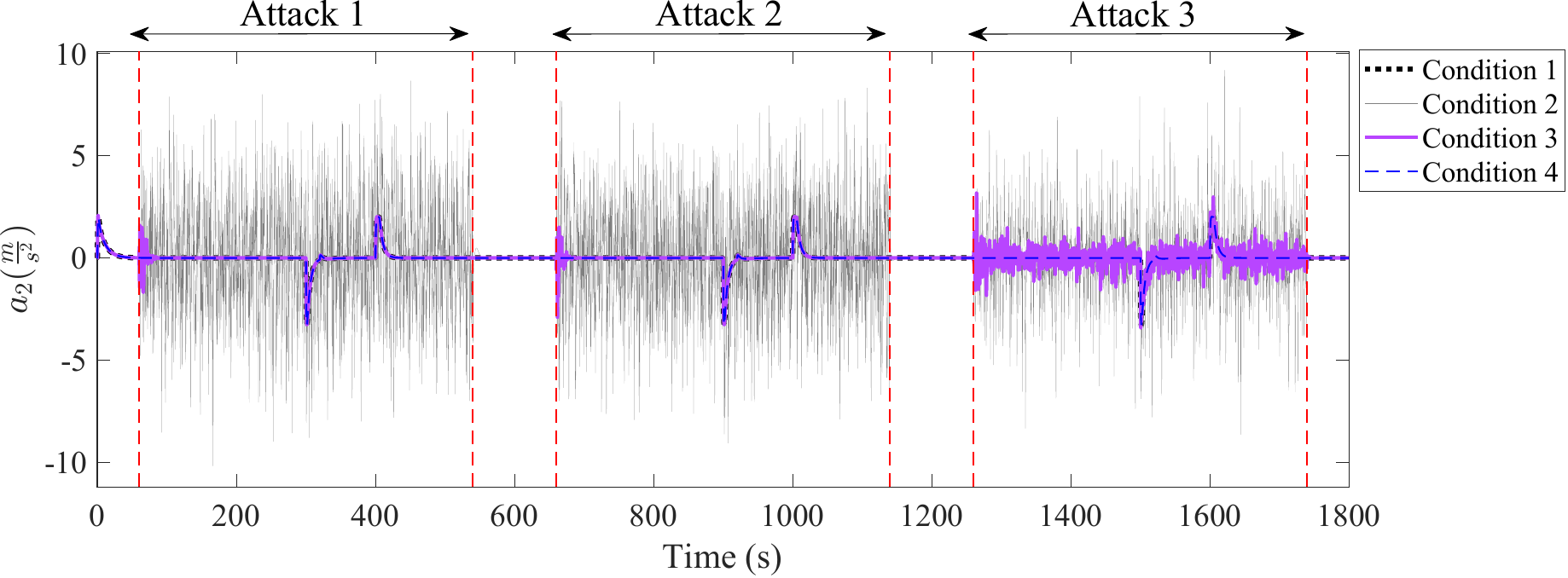}}\hspace{0 cm}
    \subfloat[]{\includegraphics[width =  1\linewidth]{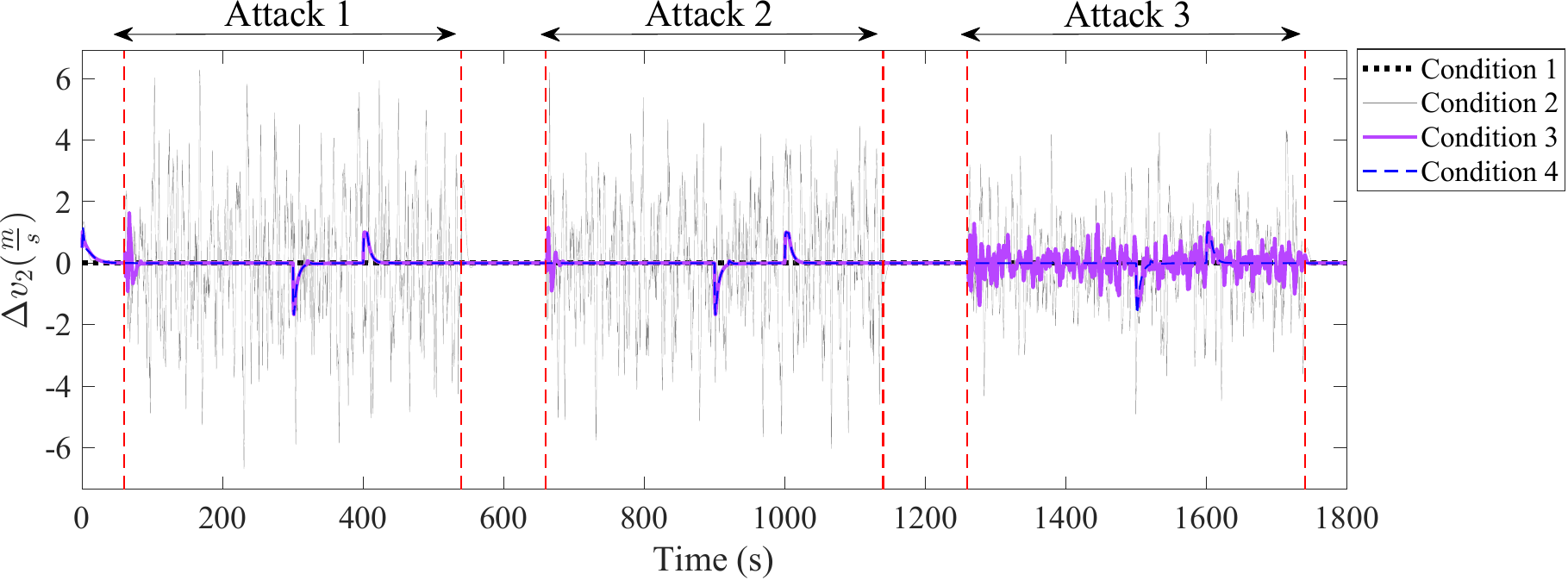}}\hspace{0 cm}

	\caption{The platooning vehicles control performance under the designed scenario including braking: a) Spacing error ($e_2$), b) Velocity ($v_2$), c) Acceleration ($a_2$), d) Relative velocity ($\Delta v_2$).}
	\label{fig:results_states_Comparision_braking}
    \vspace{\VDis}
\end{figure}

For steady state platooning as investigated in Section~\ref{subsec:results}, our method (Condition 4) performed excellently. Here we explore attacks in the more critical condition where the lead vehicle is suddenly decelerating by adding three extreme braking events to the scenario in Section~\ref{subsec:Experiment_scenario} at $t_2$, $t_5$, and $t_8$ to reach and stay on the speed of 30 $\frac{m}{s}$ for 100 $s$, and then accelerate to reach the speed of 50.4 $\frac{m}{s}$. As shown in TABLE~\ref{tab:Safety_Braking}, the braking maneuver drastically affected safety in Condition 2 as NC increased from 5 to 10 for Critical Safety attack and from 0 to 1 for Very Uncomfortable attack. From Fig.~\ref{fig:results_states_Comparision_braking}, we have the same conclusion as Fig.~\ref{fig:results_states_Comparision} which shows that Condition 4 could successfully reject the impact of attacks with no attack-related state excitation. However, Fig.~\ref{fig:results_states_Comparision_braking}(c) shows high excitation of the acceleration due to braking, which considerably affects comfort. From TABLE~\ref{tab:Comfort_Braking} and TABLE~\ref{tab:Safety_Braking}, one can see that MSDV$_x$, RC, and RMS$_{e_2}$ for Conditions 1 and 4 are similar across all severity levels, demonstrating the effectiveness of our method in rejecting the attacks' impact without any loss of comfort and security also during extreme braking.
\begin{table}[tb]
	\caption{Objective assessment of safety for braking maneuver based on Number of Collisions (NC) and RMS of the spacing error in meters (RMS$_{e_2}$).}
	\label{tab:Safety_Braking}
	\centering
	\resizebox{\linewidth}{!}{

\input{Safety_Braking}}
    \vspace{\VDis}
\end{table}
\begin{table}[tb]
	\caption{Objective assessment of comfort for braking maneuver based on Motion Sickness (MSDV$_x$) and Ride Comfort (RC) indexes.}
	\label{tab:Comfort_Braking}
	\centering
	\resizebox{\linewidth}{!}{

\input{Comfort_Braking}}
    \vspace{\VDis}
\end{table}

\vspace{-0.3 cm}
\subsection{Results: Stepwise FDI attack}
\begin{figure}[t]
	\centering
	\includegraphics[width= 1\linewidth]{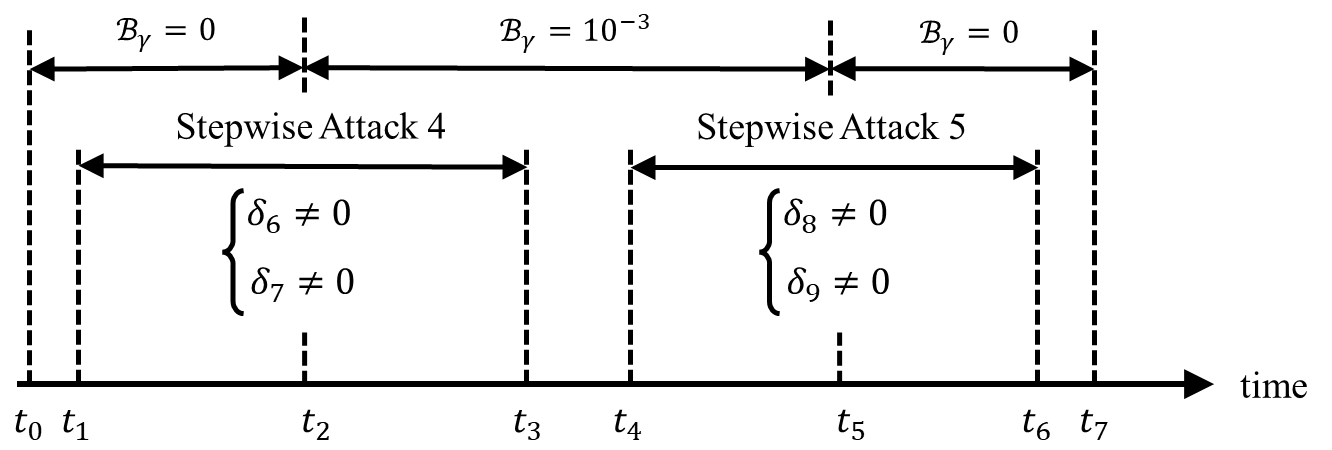}
	\caption{Designed scenario for stepwise FDI attacks.}
	\label{fig:Timing_Scenario_constantAttack}
    \vspace{\VDis}
\end{figure}

Referring to Remark~\ref{remark:eigA1}, any attack resulting in error convergence during attack occurrence may degrade the excitability of the residuals of the observer or the classification ratio. To investigate the excitability impact, we study steady state platooning using a continuous stepwise FDI attack, and a repeatedly on-off switching stepwise FDI attack \cite{taylor2022safety} with the Very Uncomfortable magnitude. 
In each of the attacks, two of the sensors are hacked at the same time, where the hacked sensors are different from one attack to the other. According to Fig.~\ref{fig:Timing_Scenario_constantAttack}, for one part of the simulation (from $t_2$ to $t_5$), sensor noise $\gamma_i$ is nonzero as a realistic condition and is i.i.d. uniformly distributed random vectors normalized to satisfy noise bounds $\mathcal{B}_{\gamma} = 10^{-3}$. For the rest of the simulation, we set $\mathcal{B}_{\gamma} = 0$. In addition, in Attack 4, sensors $y_6$ and $y_7$ are under continuous stepwise FDI attack between $t_1$ and $t_3$. In Attack 5 which is between $t_4$ and $t_6$, sensors $y_8$ and $y_9$ are under a repeatedly on-off switching stepwise FDI attack, which is activated by the activation logic in \eqref{eq:attack3}.
We performed a 20 minutes simulation where $t_0 = 0$s, $t_1 = 60$s, $t_2 = 300$s, $t_3 = 540$s, $t_4 = 660$s, $t_5 = 900$s, $t_6 = 1140$s, and $t_7 = 1200$s.

\begin{figure}[tb]
		\centering
    \subfloat[]{\includegraphics[width = 1\linewidth, height = 0.35\linewidth]{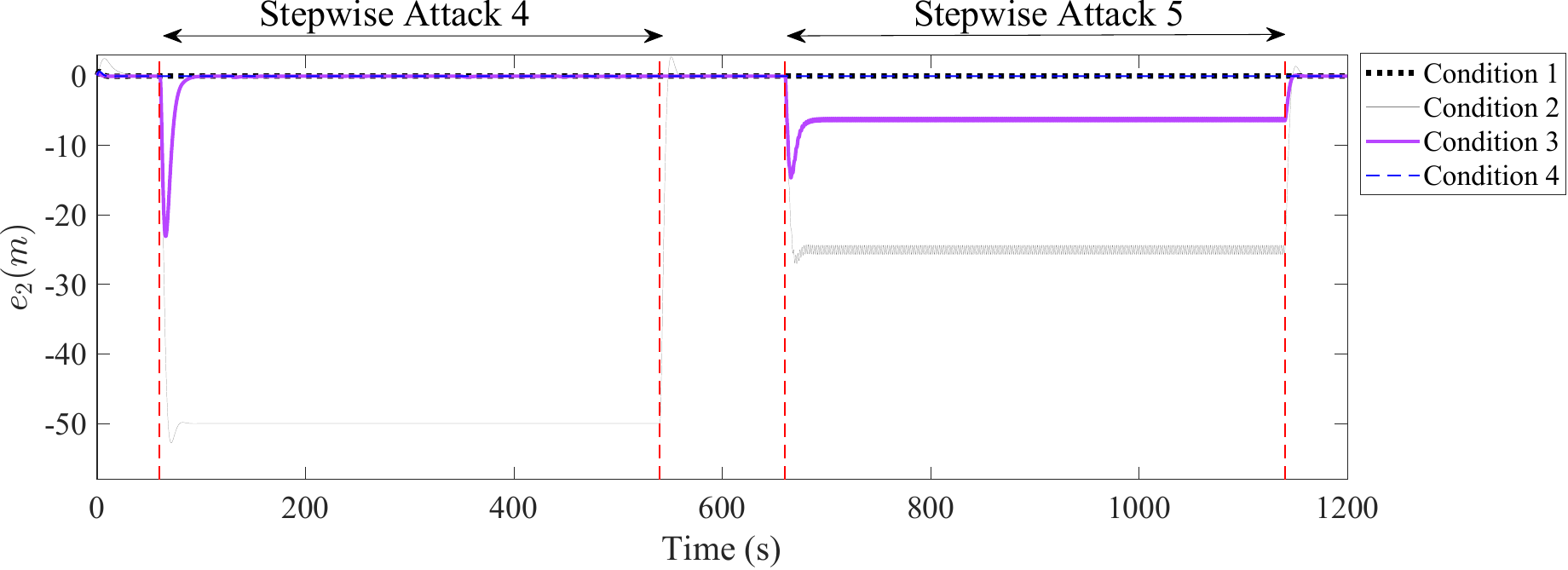}}\vspace{-0.1 cm} 
    \subfloat[]{\includegraphics[width =  1\linewidth, height = 0.35\linewidth]{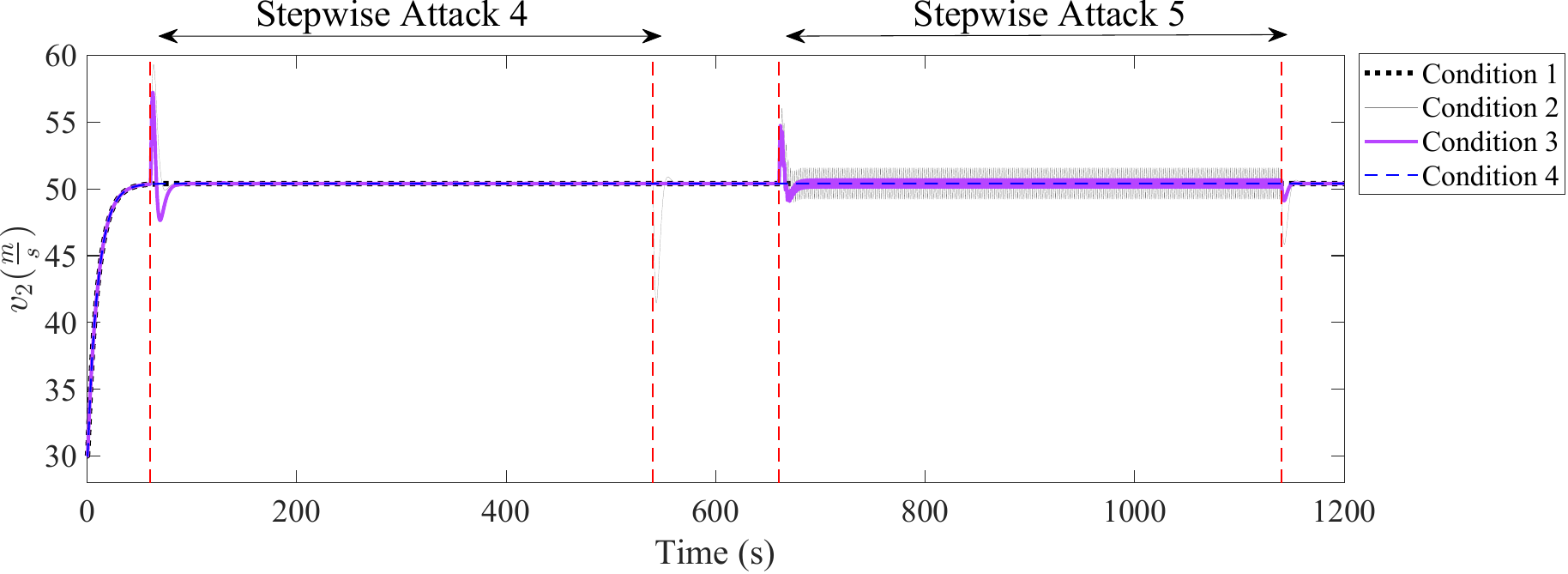}}\vspace{-0.1 cm}
    \subfloat[]{\includegraphics[width =  1\linewidth, height = 0.35\linewidth]{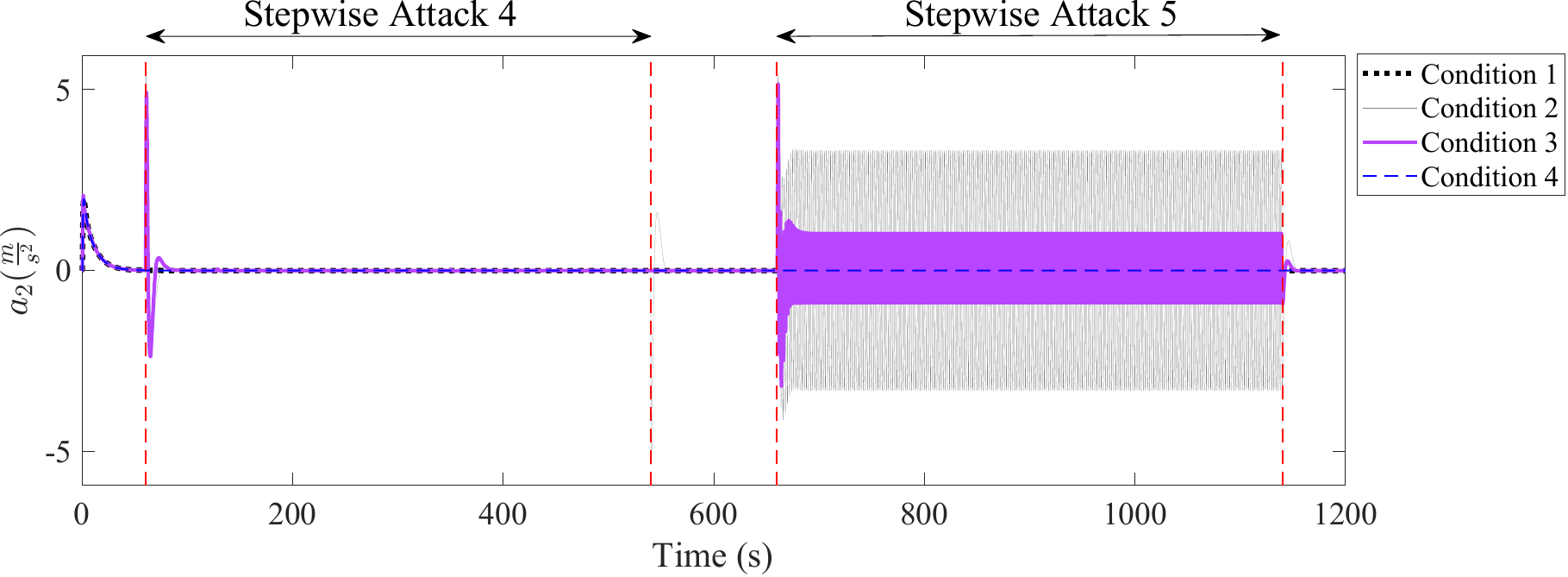}}\vspace{-0.1 cm}
    \subfloat[]{\includegraphics[width =  1\linewidth, height = 0.35\linewidth]{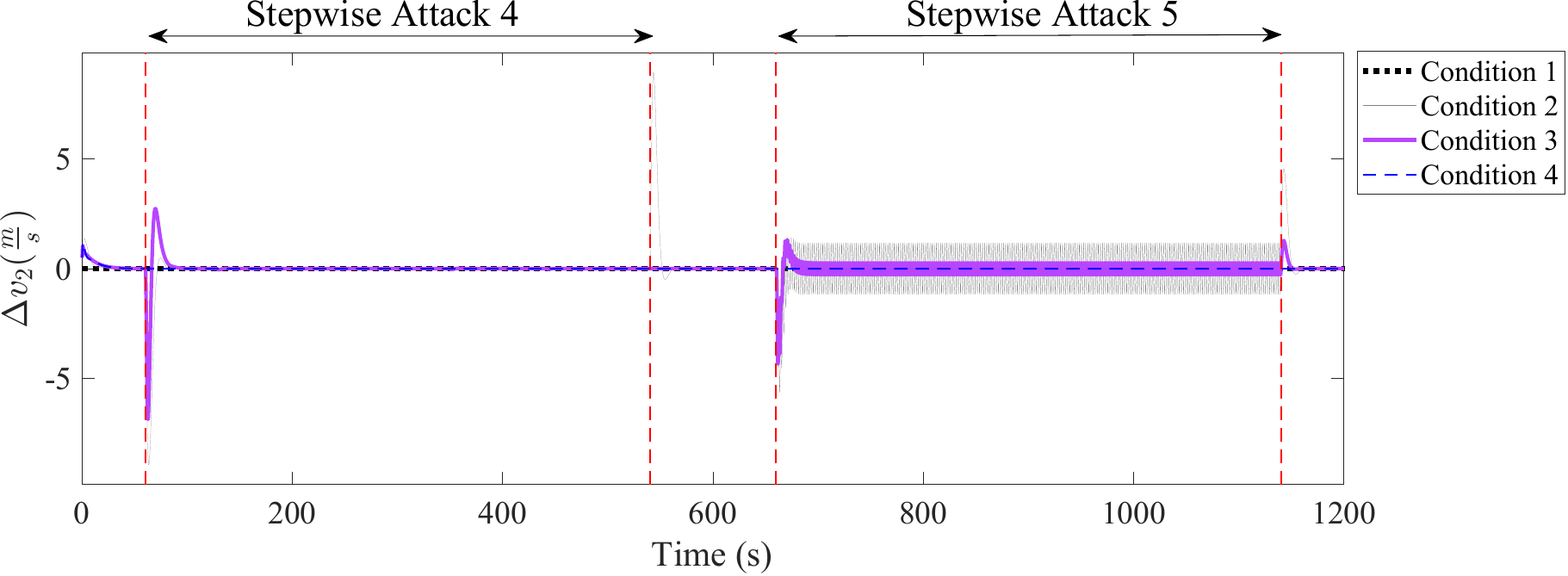}}\vspace{-0.1 cm} 
	\caption{Control performance under stepwise FDI attack: a) Spacing error ($e_2$), b) Velocity ($v_2$), c) Acceleration ($a_2$), d) Relative velocity ($\Delta v_2$).}
	\label{fig:results_states_Comparision_constant}
    \vspace{-0.4 cm}
\end{figure}

As above, we compare four conditions in the presence of stepwise FDI attacks.
The states of Vehicle 2 in the platoon are depicted in Fig.~\ref{fig:results_states_Comparision_constant}. Based on these states, the distributed CACC scheme without any security approach (Condition 2) shows the highest vulnerability to sensor attacks compared to Condition 3 and Condition 4. 
On the other hand, in Fig.~\ref{fig:results_states_Comparision_constant}, Zhao's method~\cite{zhao2021resilient} (Condition 3) presents asymptotic stability of the estimation error against Attack 4, while at the beginning of the attack, namely at $t_1$, the system states are excited for a short time. Continuously in Attack 5, Zhao's method~\cite{zhao2021resilient} (Condition 3) fails to secure the system, and Attack 5 successfully disturbs the system states due to the states' excitations in every two seconds defined in \eqref{eq:attack3}. 
However, based on Fig.~\ref{fig:results_states_Comparision_constant}, our method (Condition 4) could successfully secure the safety of vehicle platooning control by completely rejecting the impact of the two attacks. 

\vspace{-0.2cm}
\section{Conclusion and future work}
This article presents a novel optimal-coupling-observer-based framework for a homogeneous CAV platoon under bounded attacks. 
The framework includes three main components: 1) sensor set design, 2) excitation mechanism, and 3) estimation mechanism, in which all components work simultaneously together to gradually zero out the estimation error by identifying and using the most reliable observer to estimate the system states. Due to the coupling of each observer with the most reliable observer in the framework and the nonlinearity of the observer, the overall error system results in an LTV system with a complicated problem of designing the observer parameters, in which we used an LMI approach to design the parameters with guaranteed global asymptotic stability.
The results show that the framework guarantees the string stability of platoons in the presence of bounded attacks. 

Further developments include integrating the lateral behaviour, besides the longitudinal behaviour of platooning vehicles, testing in more complex scenarios such as mixed traffic flows or highly dynamic environments. From a methodology point of view, the observer structure may be regulated with a new design to achieve maximum detectability for maximum attack tolerability with minimum needed observers.
In addition, to validate the secured comfort, experiments including human participants should be considered to confirm motion comfort and perceived safety with our framework in a real environment.
%
\vspace{-0.2 cm}
\bibliographystyle{IEEEtran}
\bibliography{IEEEabrv,mybibfile}



\vspace{-0.6cm}
\begin{IEEEbiography}[{\includegraphics[width=1in,height=1.25in,clip,keepaspectratio]{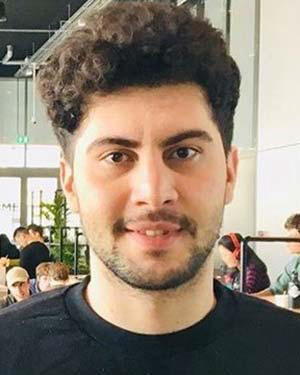}}]{Farzam Tajdari} received a Ph.D. degree from the School of Engineering, Aalto University, in 2023, and a second Ph.D. degree in mechatronic design engineering from the Delft University of Technology, in 2023. Since 2024, he has been a Postdoc Researcher with the Faculty of Mechanical Engineering, Delft University of Technology. In 2023, he was a Postdoc with the Mechanical Engineering department at the Eindhoven University of Technology, where he is currently a Guest Postdoctoral Researcher. His research interests include control, and non-linear systems, addressing challenges in the fields of ITS, privacy of dynamic systems, and geometry processing.
\end{IEEEbiography}
\vspace{-0.9cm}
\begin{IEEEbiography}[{\includegraphics[width=1in,height=1.25in,clip,keepaspectratio]{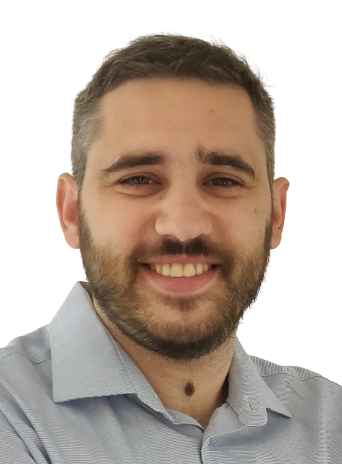}}]{Georgios Papaioannou} received the Ph.D. degree from the National Technical University of Athens (NTUA), Greece, in 2019, which received an award regarding its innovation and impact. He is currently an Assistant Professor on motion comfort in AVs at TU Delft, after conducting postdoctoral research at KTH Royal Institute of Technology in Sweden and Cranfield University in U.K. His research interests include motion comfort, seat comfort, postural stability, human body modelling, automated vehicles, motion planning, optimisation and control.
\end{IEEEbiography}
\vspace{-0.9cm}
\begin{IEEEbiography}[{\includegraphics[width=1in,height=1.25in,clip,keepaspectratio]{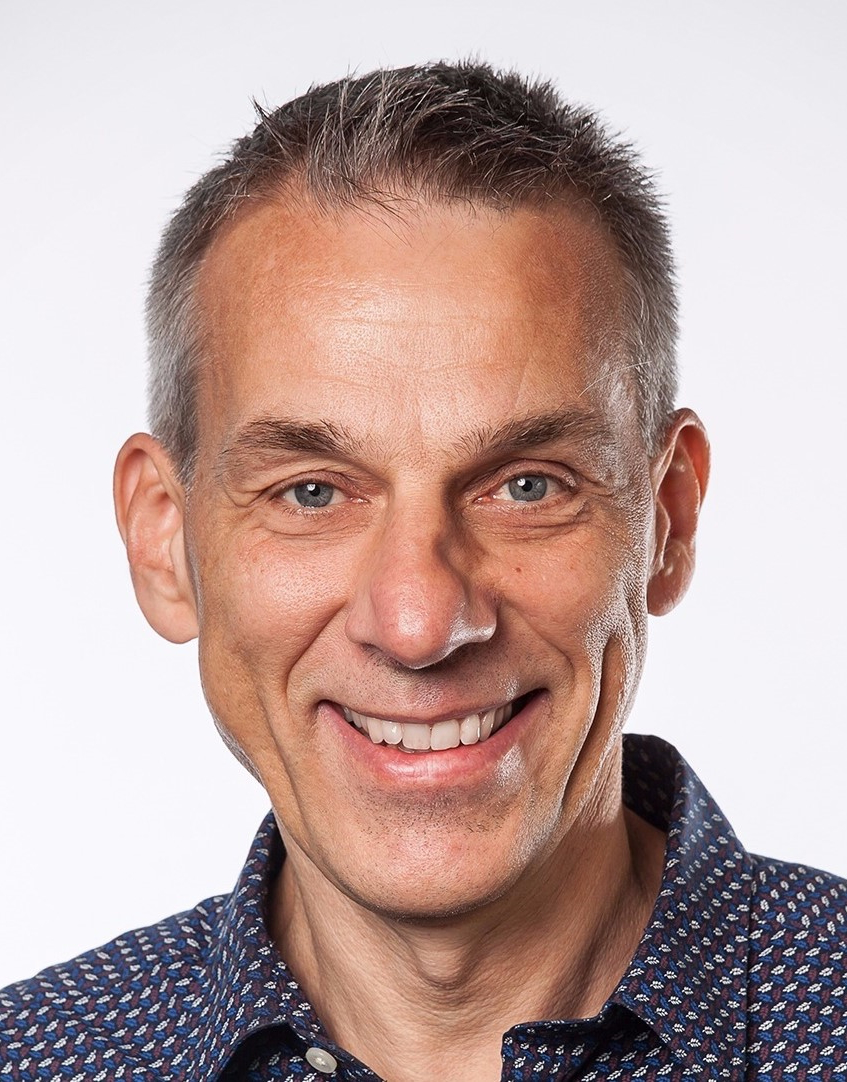}}]{Riender Happee}
received the Ph.D. degree from TU Delft, The Netherlands, in 1992. He investigated road safety and introduced biomechanical human models for impact and comfort at TNO Automotive (1992-2007). Currently, he investigates the human interaction with automated vehicles focusing on motion comfort, perceived safety and acceptance at the Delft University of Technology, the Netherlands, where he is full Professor.
\end{IEEEbiography}



\vfill

\end{document}

%% file: RMS_attack_scenario.tex
	
\begin{tabular}{c c c c c c c c c}
	\hline\hline
	\multirow{2}{*}{Severity level} & \multicolumn{2}{c}{Attack 1} & & \multicolumn{2}{c}{Attack 2}& & \multicolumn{2}{c}{Attack 3}\\ \cline{2-3} \cline{5-6} \cline{8-9}
	& $\delta_1~(m)$ & $\delta_2~(\frac{m}{s})$ & & $\delta_6~(m)$ & $\delta_7~(\frac{m}{s})$ & & $\delta_8~(m)$ & $\delta_9~(\frac{m}{s})$\\
	\hline \hline
	Critical Safety & 300 & 300 & & 300 & 300  & & 300 & 300 \\ \hline
	Very Uncomfortable & 150 & 150 & & 150 & 150  & & 150 & 150 \\ \hline
	Uncomfortable & 15 & 15 & & 15 & 15  & & 15 & 15 \\
	\hline\hline
\end{tabular}

%% file: Safety_SteadyState.tex
\begin{tabular}{c c c c c c c}

	\hline\hline
 \parbox[t]{1.8cm}{\multirow{3}{*}{Condition}} & \multicolumn{6}{c}{Severity level}\\ \cline{2-7}
 & \multicolumn{2}{c}{Critical Safety} &  \multicolumn{2}{c}{Very Uncomfortable} &  \multicolumn{2}{c}{Uncomfortable} \\  
 \addlinespace[-0.03em]\cmidrule(l){2-3}
 \cmidrule(l){4-5}
 \cmidrule(l){6-7}
 & NC & RMS$_{e_2}$ & NC & RMS$_{e_2}$ & NC & RMS$_{e_2}$ \\
\hline\hline

 \parbox[t]{1.8cm}{1 (Insecure controller ~\cite{ploeg2013lp} without attack)}& 0 & 0.028 & 0 & 0.028 & 0 & 0.028 \\ \hline
 \parbox[t]{1.8cm}{2 (Insecure controller~\cite{ploeg2013lp} with attack)}& 5 & 9.516 & 0 & 4.912 & 0 & 0.547 \\ \hline
 \parbox[t]{1.8cm}{3 (Zhao's method~\cite{zhao2021resilient} with attack)}& 0 & 0.774 & 0 & 0.579 & 0 & 0.079 \\ \hline
 \parbox[t]{1.8cm}{4 (Our method with attack)}& 0 & 0.029 & 0 & 0.029 & 0 & 0.029 \\

	\hline\hline
\end{tabular}

%% file: Comfort_SteadyState.tex
\begin{tabular}{c c c c c c c}

	\hline\hline
 \parbox[t]{1.8cm}{\multirow{3}{*}{Condition}} & \multicolumn{6}{c}{Severity level}\\ \cline{2-7}
 & \multicolumn{2}{c}{Critical Safety} &  \multicolumn{2}{c}{Very Uncomfortable} &  \multicolumn{2}{c}{Uncomfortable} \\  
 \addlinespace[-0.03em]\cmidrule(l){2-3}
 \cmidrule(l){4-5}
 \cmidrule(l){6-7}
 & MSDV$_x$ & RC & MSDV$_x$ & RC & MSDV$_x$ & RC\\
\hline\hline

 \parbox[t]{1.8cm}{1 (Insecure controller~\cite{ploeg2013lp} without attack)}& 0.0028 & 4e-5 & 0.0028 & 4e-5 & 0.0028 & 4e-5 \\ \hline
 \parbox[t]{1.8cm}{2 (Insecure controller~\cite{ploeg2013lp} with attack)}& 94.69 & 3.08 & 47.38 & 1.59 & 4.65 & 0.15 \\ \hline
 \parbox[t]{1.8cm}{3 (Zhao's method~\cite{zhao2021resilient} with attack)}& 10.41 & 0.29 & 6.57 & 0.16 & 1.01 & 0.02 \\ \hline
 \parbox[t]{1.8cm}{4 (Our method with attack)}& 0.0029 & 5e-5 & 0.0029 & 5e-5 & 0.0029 & 5e-5 \\

	\hline\hline
\end{tabular}

%% file: Safety_Braking.tex
\begin{tabular}{c c c c c c c}

	\hline\hline
 \parbox[t]{1.8cm}{\multirow{3}{*}{Condition}} & \multicolumn{6}{c}{Severity level}\\ \cline{2-7}
 & \multicolumn{2}{c}{Critical Safety} &  \multicolumn{2}{c}{Very Uncomfortable} &  \multicolumn{2}{c}{Uncomfortable} \\  
 \addlinespace[-0.03em]\cmidrule(l){2-3}
 \cmidrule(l){4-5}
 \cmidrule(l){6-7}
 & NC & RMS$_{e_2}$ & NC & RMS$_{e_2}$ & NC & RMS$_{e_2}$ \\
\hline\hline

 \parbox[t]{1.8cm}{1 (Insecure controller~\cite{ploeg2013lp} without attack)} & 0 & 0.03 & 0 & 0.03 & 0 & 0.03 \\ \hline
 \parbox[t]{1.8cm}{2 (Insecure controller~\cite{ploeg2013lp} with attack)} & 10 & 9.76 & 1 & 4.54 & 0 & 0.77 \\ \hline
 \parbox[t]{1.8cm}{3 (Zhao's method~\cite{zhao2021resilient} with attack)}& 0 & 1.15 & 0 & 0.53 & 0 & 0.10 \\ \hline
 \parbox[t]{1.8cm}{4 (Our method with attack)}& 0 & 0.05 & 0 & 0.03 & 0 & 0.03 \\

	\hline\hline
\end{tabular}

%% file: Comfort_Braking.tex
\begin{tabular}{c c c c c c c}

	\hline\hline
 \parbox[t]{1.8cm}{\multirow{3}{*}{Condition}} & \multicolumn{6}{c}{Severity level}\\ \cline{2-7}
 & \multicolumn{2}{c}{Critical Safety} &  \multicolumn{2}{c}{Very Uncomfortable} &  \multicolumn{2}{c}{Uncomfortable} \\  
 \addlinespace[-0.03em]\cmidrule(l){2-3}
 \cmidrule(l){4-5}
 \cmidrule(l){6-7}
 & MSDV$_x$ & RC & MSDV$_x$ & RC & MSDV$_x$ & RC\\
\hline\hline

 \parbox[t]{1.8cm}{1 (Insecure controller~\cite{ploeg2013lp} without attack)}& 3.81 & 0.03 & 3.81 & 0.03 & 3.81 & 0.03 \\ \hline
 \parbox[t]{1.8cm}{2 (Insecure controller~\cite{ploeg2013lp} with attack)}   & 97.41& 3.17 & 48.21 & 1.59 & 5.63 & 0.16 \\ \hline
 \parbox[t]{1.8cm}{3 (Zhao's method~\cite{zhao2021resilient} with attack)}      & 13.13& 0.27 & 7.09 & 0.15 & 3.94 & 0.04 \\ \hline
 \parbox[t]{1.8cm}{4 (Our method with attack)}           & 3.81 & 0.03 & 3.81 & 0.03 & 3.81 & 0.03 \\

	\hline\hline
\end{tabular}

%% file: Main_Final.bbl
\begin{thebibliography}{10}
\providecommand{\url}[1]{#1}
\csname url@samestyle\endcsname
\providecommand{\newblock}{\relax}
\providecommand{\bibinfo}[2]{#2}
\providecommand{\BIBentrySTDinterwordspacing}{\spaceskip=0pt\relax}
\providecommand{\BIBentryALTinterwordstretchfactor}{4}
\providecommand{\BIBentryALTinterwordspacing}{\spaceskip=\fontdimen2\font plus
\BIBentryALTinterwordstretchfactor\fontdimen3\font minus
  \fontdimen4\font\relax}
\providecommand{\BIBforeignlanguage}[2]{{%
\expandafter\ifx\csname l@#1\endcsname\relax
\typeout{** WARNING: IEEEtran.bst: No hyphenation pattern has been}%
\typeout{** loaded for the language `#1'. Using the pattern for}%
\typeout{** the default language instead.}%
\else
\language=\csname l@#1\endcsname
\fi
#2}}
\providecommand{\BIBdecl}{\relax}
\BIBdecl

\bibitem{lu2014connected}
N.~Lu, N.~Cheng, N.~Zhang, X.~Shen, and J.~W. Mark, ``Connected vehicles:
  Solutions and challenges,'' \emph{IEEE internet of things journal}, vol.~1,
  no.~4, pp. 289--299, 2014.

\bibitem{guerrero2015integration}
J.~A. Guerrero-Ibanez, S.~Zeadally, and J.~Contreras-Castillo, ``Integration
  challenges of intelligent transportation systems with connected vehicle,
  cloud computing, and internet of things technologies,'' \emph{IEEE Wireless
  Communications}, vol.~22, no.~6, pp. 122--128, 2015.

\bibitem{khalil2022connected}
A.~Khalil, K.~F. Aljanaideh, and M.~Al~Janaideh, ``On connected autonomous
  vehicles with unknown human driven vehicles effects using transmissibility
  operators,'' \emph{IEEE Transactions on Automation Science and Engineering},
  2022.

\bibitem{tajdari2020feedback}
F.~Tajdari, C.~Roncoli, and M.~Papageorgiou, ``Feedback-based ramp metering and
  lane-changing control with connected and automated vehicles,'' \emph{IEEE
  Transactions on Intelligent Transportation Systems}, vol.~23, no.~2, pp.
  939--951, 2020.

\bibitem{tajdari2021adaptive}
F.~Tajdari and C.~Roncoli, ``Adaptive traffic control at motorway bottlenecks
  with time-varying fundamental diagram,'' \emph{IFAC-PapersOnLine}, vol.~54,
  no.~2, pp. 271--277, 2021.

\bibitem{tajdari2019integrated}
F.~Tajdari, C.~Roncoli, N.~Bekiaris-Liberis, and M.~Papageorgiou, ``Integrated
  ramp metering and lane-changing feedback control at motorway bottlenecks,''
  in \emph{2019 18th European Control Conference (ECC)}.\hskip 1em plus 0.5em
  minus 0.4em\relax IEEE, 2019, pp. 3179--3184.

\bibitem{li2017dynamical}
S.~E. Li, Y.~Zheng, K.~Li, Y.~Wu, J.~K. Hedrick, F.~Gao, and H.~Zhang,
  ``Dynamical modeling and distributed control of connected and automated
  vehicles: Challenges and opportunities,'' \emph{IEEE Intelligent
  Transportation Systems Magazine}, vol.~9, no.~3, pp. 46--58, 2017.

\bibitem{zheng2017platooning}
Y.~Zheng, S.~E. Li, K.~Li, and W.~Ren, ``Platooning of connected vehicles with
  undirected topologies: Robustness analysis and distributed h-infinity
  controller synthesis,'' \emph{IEEE Transactions on Intelligent Transportation
  Systems}, vol.~19, no.~5, pp. 1353--1364, 2017.

\bibitem{zhou2022robust}
J.~Zhou, D.~Tian, Z.~Sheng, X.~Duan, G.~Qu, D.~Zhao, D.~Cao, and X.~Shen,
  ``Robust min-max model predictive vehicle platooning with causal disturbance
  feedback,'' \emph{IEEE Transactions on Intelligent Transportation Systems},
  vol.~23, no.~9, pp. 15\,878--15\,897, 2022.

\bibitem{li2023adaptive}
Y.~Li, Y.~Zhao, and S.~Tong, ``Adaptive fuzzy control for heterogeneous
  vehicular platoon systems with collision avoidance and connectivity
  preservation,'' \emph{IEEE Transactions on Fuzzy Systems}, 2023.

\bibitem{tajdari2023online}
F.~Tajdari and C.~Roncoli, ``Online set-point estimation for feedback-based
  traffic control applications,'' \emph{IEEE Transactions on Intelligent
  Transportation Systems}, vol.~24, no.~10, pp. 10\,830--10\,842, 2023.

\bibitem{diels2016self}
C.~Diels and J.~E. Bos, ``Self-driving carsickness,'' \emph{Applied
  Ergonomics}, vol.~53, pp. 374--382, 2016.

\bibitem{papaioannou2025occupants}
G.~Papaioannou, C.~Shen, M.~Rothh{\"a}mel, and R.~Happee, ``Occupants' comfort:
  what about human body dynamics in road and rail vehicles?'' \emph{Vehicle
  System Dynamics}, pp. 1--59, 2025.

\bibitem{aledhari2023motion}
M.~Aledhari, M.~Rahouti, J.~Qadir, B.~Qolomany, M.~Guizani, and A.~Al-Fuqaha,
  ``Motion comfort optimization for autonomous vehicles: Concepts, methods, and
  techniques,'' \emph{IEEE Internet of Things Journal}, vol.~11, no.~1, pp.
  378--402, 2023.

\bibitem{he2024new}
X.~He, R.~Happee, and M.~Wang, ``A new computational perceived risk model for
  automated vehicles based on potential collision avoidance difficulty
  (pcad),'' \emph{Transportation Research Part C: Emerging Technologies}, vol.
  166, p. 104751, 2024.

\bibitem{he2022modelling}
X.~He, J.~Stapel, M.~Wang, and R.~Happee, ``Modelling perceived risk and trust
  in driving automation reacting to merging and braking vehicles,''
  \emph{Transportation research part F: traffic psychology and behaviour},
  vol.~86, pp. 178--195, 2022.

\bibitem{lian2022fuzzy}
Z.~Lian, P.~Shi, C.-C. Lim, and X.~Yuan, ``Fuzzy-model-based lateral control
  for networked autonomous vehicle systems under hybrid cyber-attacks,''
  \emph{IEEE Transactions on Cybernetics}, vol.~53, no.~4, pp. 2600--2609,
  2022.

\bibitem{bian2021fuel}
Y.~Bian, C.~Du, M.~Hu, S.~E. Li, H.~Liu, and C.~Li, ``Fuel economy optimization
  for platooning vehicle swarms via distributed economic model predictive
  control,'' \emph{IEEE Transactions on Automation Science and Engineering},
  vol.~19, no.~4, pp. 2711--2723, 2021.

\bibitem{li2022adaptive}
Y.~Li, Y.~Zhao, W.~Liu, and J.~Hu, ``Adaptive fuzzy predefined-time control for
  third-order heterogeneous vehicular platoon systems with dead-zone,''
  \emph{IEEE Transactions on Industrial Informatics}, 2022.

\bibitem{petit2014potential}
J.~Petit and S.~E. Shladover, ``Potential cyberattacks on automated vehicles,''
  \emph{IEEE Transactions on Intelligent transportation systems}, vol.~16,
  no.~2, pp. 546--556, 2014.

\bibitem{deng2021deep}
Y.~Deng, T.~Zhang, G.~Lou, X.~Zheng, J.~Jin, and Q.-L. Han, ``Deep
  learning-based autonomous driving systems: A survey of attacks and
  defenses,'' \emph{IEEE Transactions on Industrial Informatics}, vol.~17,
  no.~12, pp. 7897--7912, 2021.

\bibitem{adhikari2021security}
M.~Adhikari, A.~Munusamy, A.~Hazra, V.~G. Menon, V.~Anavangot, and D.~Puthal,
  ``Security in edge-centric intelligent internet of vehicles: Issues and
  remedies,'' \emph{IEEE Consumer Electronics Magazine}, vol.~11, no.~6, pp.
  24--31, 2021.

\bibitem{nikitas2022deceitful}
A.~Nikitas, S.~Parkinson, and M.~Vallati, ``The deceitful connected and
  autonomous vehicle: Defining the concept, contextualising its dimensions and
  proposing mitigation policies,'' \emph{Transport policy}, vol. 122, pp.
  1--10, 2022.

\bibitem{nordhoff2023driver}
S.~Nordhoff, J.~Stapel, X.~He, A.~Gentner, and R.~Happee, ``Do driver’s
  characteristics, system performance, perceived safety, and trust influence
  how drivers use partial automation? a structural equation modelling
  analysis,'' \emph{Frontiers in Psychology}, vol.~14, p. 1125031, 2023.

\bibitem{nordhoff2020using}
S.~Nordhoff, T.~Louw, S.~Innamaa, E.~Lehtonen, A.~Beuster, G.~Torrao,
  A.~Bjorvatn, T.~Kessel, F.~Malin, R.~Happee \emph{et~al.}, ``Using the utaut2
  model to explain public acceptance of conditionally automated (l3) cars: A
  questionnaire study among 9,118 car drivers from eight european countries,''
  \emph{Transportation research part F: traffic psychology and behaviour},
  vol.~74, pp. 280--297, 2020.

\bibitem{li2023bumpless}
Y.~Li and S.~Tong, ``Bumpless transfer distributed adaptive backstepping
  control of nonlinear multi-agent systems with circular filtering under dos
  attacks,'' \emph{Automatica}, vol. 157, p. 111250, 2023.

\bibitem{zhao2021resilient}
Y.~Zhao, Z.~Liu, and W.~S. Wong, ``Resilient platoon control of vehicular cyber
  physical systems under dos attacks and multiple disturbances,'' \emph{IEEE
  Transactions on Intelligent Transportation Systems}, vol.~23, no.~8, pp.
  10\,945--10\,956, 2021.

\bibitem{biroon2021false}
R.~A. Biroon, Z.~A. Biron, and P.~Pisu, ``False data injection attack in a
  platoon of cacc: real-time detection and isolation with a pde approach,''
  \emph{IEEE transactions on intelligent transportation systems}, vol.~23,
  no.~7, pp. 8692--8703, 2021.

\bibitem{xie2022distributed}
M.~Xie, D.~Ding, X.~Ge, Q.-L. Han, H.~Dong, and Y.~Song, ``Distributed
  platooning control of automated vehicles subject to replay attacks based on
  proportional integral observers,'' \emph{IEEE/CAA Journal of Automatica
  Sinica}, 2022.

\bibitem{ju2020deception}
Z.~Ju, H.~Zhang, and Y.~Tan, ``Deception attack detection and estimation for a
  local vehicle in vehicle platooning based on a modified ufir estimator,''
  \emph{IEEE Internet of Things Journal}, vol.~7, no.~5, pp. 3693--3705, 2020.

\bibitem{yang2022resource}
F.~Yang, Z.~Gu, L.~Hua, and S.~Yan, ``A resource-aware control approach to
  vehicle platoons under false data injection attacks,'' \emph{ISA
  transactions}, vol. 131, pp. 367--376, 2022.

\bibitem{zhou2022attack}
Z.~Zhou, F.~Zhu, D.~Xu, S.~Guo, and Y.~Zhao, ``Attack resilient control for
  vehicle platoon system with full states constraint under actuator faulty
  scenario,'' \emph{Applied Mathematics and Computation}, vol. 419, p. 126874,
  2022.

\bibitem{chong2015observability}
M.~S. Chong, M.~Wakaiki, and J.~P. Hespanha, ``Observability of linear systems
  under adversarial attacks,'' in \emph{2015 American Control Conference
  (ACC)}.\hskip 1em plus 0.5em minus 0.4em\relax IEEE, 2015, pp. 2439--2444.

\bibitem{fawzi2014secure}
H.~Fawzi, P.~Tabuada, and S.~Diggavi, ``Secure estimation and control for
  cyber-physical systems under adversarial attacks,'' \emph{IEEE Transactions
  on Automatic control}, vol.~59, no.~6, pp. 1454--1467, 2014.

\bibitem{shoukry2014event}
Y.~Shoukry and P.~Tabuada, ``Event-triggered projected luenberger observer for
  linear systems under sparse sensor attacks,'' in \emph{53rd IEEE Conference
  on Decision and Control}.\hskip 1em plus 0.5em minus 0.4em\relax IEEE, 2014,
  pp. 3548--3553.

\bibitem{hassani2009multiple}
V.~Hassani, A.~P. Aguiar, M.~Athans, and A.~M. Pascoal, ``Multiple model
  adaptive estimation and model identification usign a minimum energy
  criterion,'' in \emph{2009 American Control Conference}.\hskip 1em plus 0.5em
  minus 0.4em\relax IEEE, 2009, pp. 518--523.

\bibitem{elbanhawi2015passenger}
M.~Elbanhawi, M.~Simic, and R.~Jazar, ``In the passenger seat: investigating
  ride comfort measures in autonomous cars,'' \emph{IEEE Intelligent
  transportation systems magazine}, vol.~7, no.~3, pp. 4--17, 2015.

\bibitem{htike2021fundamentals}
Z.~Htike, G.~Papaioannou, E.~Siampis, E.~Velenis, and S.~Longo, ``Fundamentals
  of motion planning for mitigating motion sickness in automated vehicles,''
  \emph{IEEE Transactions on Vehicular Technology}, vol.~71, no.~3, pp.
  2375--2384, 2021.

\bibitem{jain2023optimal}
V.~Jain, S.~S. Kumar, G.~Papaioannou, R.~Happee, and B.~Shyrokau, ``Optimal
  trajectory planning for mitigated motion sickness: Simulator study
  assessment,'' \emph{IEEE Transactions on Intelligent Transportation Systems},
  vol.~24, no.~10, pp. 10\,653--10\,664, 2023.

\bibitem{tajdari2019fuzzy}
F.~Tajdari, A.~Ghaffari, A.~Khodayari, A.~Kamali, N.~Zhilakzadeh, and
  N.~Ebrahimi, ``Fuzzy control of anticipation and evaluation behaviour in real
  traffic flow,'' in \emph{2019 7th International Conference on Robotics and
  Mechatronics (ICRoM)}.\hskip 1em plus 0.5em minus 0.4em\relax IEEE, 2019, pp.
  248--253.

\bibitem{tajdari2021simultaneous}
F.~Tajdari, A.~Golgouneh, A.~Ghaffari, A.~Khodayari, A.~Kamali, and
  N.~Hosseinkhani, ``Simultaneous intelligent anticipation and control of
  follower vehicle observing exiting lane changer,'' \emph{IEEE Transactions on
  Vehicular Technology}, vol.~70, no.~9, pp. 8567--8577, 2021.

\bibitem{tajdari2020intelligent}
F.~Tajdari, N.~E. Toulkani, and M.~Nourimand, ``Intelligent architecture for
  car-following behaviour observing lane-changer: Modeling and control,'' in
  \emph{2020 10th International Conference on Computer and Knowledge
  Engineering (ICCKE)}.\hskip 1em plus 0.5em minus 0.4em\relax IEEE, 2020, pp.
  579--584.

\bibitem{tajdari2022flow}
F.~Tajdari, H.~Ramezanian, S.~Paydarfar, A.~Lashgari, and S.~Maghrebi, ``Flow
  metering and lane-changing optimal control with ramp-metering saturation,''
  in \emph{2022 CPSSI 4th International Symposium on Real-Time and Embedded
  Systems and Technologies (RTEST)}.\hskip 1em plus 0.5em minus 0.4em\relax
  IEEE, 2022, pp. 1--6.

\bibitem{papaioannou2022multi}
G.~Papaioannou, Z.~Htike, C.~Lin, E.~Siampis, S.~Longo, and E.~Velenis,
  ``Multi-criteria evaluation for sorting motion planner alternatives,''
  \emph{Sensors}, vol.~22, no.~14, p. 5177, 2022.

\bibitem{limbasiya2022systematic}
T.~Limbasiya, K.~Z. Teng, S.~Chattopadhyay, and J.~Zhou, ``A systematic survey
  of attack detection and prevention in connected and autonomous vehicles,''
  \emph{Vehicular Communications}, vol.~37, p. 100515, 2022.

\bibitem{rajamani2002semi}
R.~Rajamani and C.~Zhu, ``Semi-autonomous adaptive cruise control systems,''
  \emph{IEEE Transactions on Vehicular Technology}, vol.~51, no.~5, pp.
  1186--1192, 2002.

\bibitem{linkov2019human}
V.~Linkov, P.~Z{\'a}me{\v{c}}n{\'\i}k, D.~Havl{\'\i}{\v{c}}kov{\'a}, and C.-W.
  Pai, ``Human factors in the cybersecurity of autonomous vehicles: Trends in
  current research,'' \emph{Frontiers in psychology}, vol.~10, p. 995, 2019.

\bibitem{wang2020modeling}
P.~Wang, X.~Wu, and X.~He, ``Modeling and analyzing cyberattack effects on
  connected automated vehicular platoons,'' \emph{Transportation research part
  C: emerging technologies}, vol. 115, p. 102625, 2020.

\bibitem{shet2019performance}
R.~A. Shet and F.~Schewe, ``Performance evaluation of cruise controls and their
  impact on passenger comfort in autonomous vehicle platoons,'' in \emph{2019
  ieee 89th vehicular technology conference (vtc2019-spring)}.\hskip 1em plus
  0.5em minus 0.4em\relax IEEE, 2019, pp. 1--7.

\bibitem{kuang2024research}
J.~Kuang, G.~Tan, X.~Guo, X.~Pei, and D.~Peng, ``Research of obstacle vehicles
  avoidance for automated heavy vehicle platoon by switching the formation,''
  \emph{IET Intelligent Transport Systems}, vol.~18, no.~4, pp. 630--644, 2024.

\bibitem{liu2023efficient}
J.~Liu, Z.~Wang, and L.~Zhang, ``Efficient eco-driving control for ev platoons
  in mixed urban traffic scenarios considering regenerative braking,''
  \emph{IEEE Transactions on Transportation Electrification}, 2023.

\bibitem{oncu2014cooperative}
S.~{\"O}nc{\"u}, J.~Ploeg, N.~Van~de Wouw, and H.~Nijmeijer, ``Cooperative
  adaptive cruise control: Network-aware analysis of string stability,''
  \emph{IEEE Transactions on Intelligent Transportation Systems}, vol.~15,
  no.~4, pp. 1527--1537, 2014.

\bibitem{wolf2004security}
M.~Wolf, A.~Weimerskirch, and C.~Paar, ``Security in automotive bus systems,''
  in \emph{Workshop on Embedded Security in Cars}.\hskip 1em plus 0.5em minus
  0.4em\relax Bochum, 2004, pp. 1--13.

\bibitem{zhou2023robust}
A.~Zhou, J.~Wang, and S.~Peeta, ``Robust control strategy for platoon of
  connected and autonomous vehicles considering falsified information injected
  through communication links,'' \emph{Journal of Intelligent Transportation
  Systems}, vol.~27, no.~6, pp. 735--751, 2023.

\bibitem{jin2019adaptive}
X.~Jin, W.~M. Haddad, Z.-P. Jiang, A.~Kanellopoulos, and K.~G. Vamvoudakis,
  ``An adaptive learning and control architecture for mitigating sensor and
  actuator attacks in connected autonomous vehicle platoons,''
  \emph{International Journal of Adaptive Control and Signal Processing},
  vol.~33, no.~12, pp. 1788--1802, 2019.

\bibitem{yadegar2019output}
M.~Yadegar, N.~Meskin, and W.~M. Haddad, ``An output-feedback adaptive control
  architecture for mitigating actuator attacks in cyber-physical systems,''
  \emph{International Journal of Adaptive Control and Signal Processing},
  vol.~33, no.~6, pp. 943--955, 2019.

\bibitem{jahanshahi2018attack}
N.~Jahanshahi and R.~M. Ferrari, ``Attack detection and estimation in
  cooperative vehicles platoons: A sliding mode observer approach,''
  \emph{IFAC-PapersOnLine}, vol.~51, no.~23, pp. 212--217, 2018.

\bibitem{li2023secure}
Z.~Li, M.~U.~B. Niazi, C.~Liu, Y.~Mo, and K.~H. Johansson, ``Secure state
  estimation against sparse attacks on a time-varying set of sensors,''
  \emph{IFAC-PapersOnLine}, vol.~56, no.~2, pp. 270--275, 2023.

\bibitem{petrillo2017collaborative}
A.~Petrillo, A.~Pescap{\'e}, and S.~Santini, ``A collaborative control strategy
  for platoons of autonomous vehicles in the presence of message falsification
  attacks,'' in \emph{2017 5th IEEE International Conference on Models and
  Technologies for Intelligent Transportation Systems (MT-ITS)}.\hskip 1em plus
  0.5em minus 0.4em\relax IEEE, 2017, pp. 110--115.

\bibitem{kremer2020state}
P.~Kremer, I.~Koley, S.~Dey, and S.~Park, ``State estimation for attack
  detection in vehicle platoon using vanet and controller model,'' in
  \emph{2020 IEEE 23rd International Conference on Intelligent Transportation
  Systems (ITSC)}.\hskip 1em plus 0.5em minus 0.4em\relax IEEE, 2020, pp. 1--8.

\bibitem{merco2018replay}
R.~Merco, Z.~A. Biron, and P.~Pisu, ``Replay attack detection in a platoon of
  connected vehicles with cooperative adaptive cruise control,'' in \emph{2018
  Annual American Control Conference (ACC)}.\hskip 1em plus 0.5em minus
  0.4em\relax IEEE, 2018, pp. 5582--5587.

\bibitem{ploeg2013lp}
J.~Ploeg, N.~Van De~Wouw, and H.~Nijmeijer, ``Lp string stability of cascaded
  systems: Application to vehicle platooning,'' \emph{IEEE Transactions on
  Control Systems Technology}, vol.~22, no.~2, pp. 786--793, 2013.

\bibitem{shen2022cooperative}
Z.~Shen, Y.~Liu, Z.~Li, and M.~H. Nabin, ``Cooperative spacing sampled control
  of vehicle platoon considering undirected topology and analog fading
  networks,'' \emph{IEEE Transactions on Intelligent Transportation Systems},
  vol.~23, no.~10, pp. 18\,478--18\,491, 2022.

\bibitem{vegamoor2021string}
V.~Vegamoor, S.~Rathinam, and S.~Darbha, ``String stability of connected
  vehicle platoons under lossy v2v communication,'' \emph{IEEE Transactions on
  Intelligent Transportation Systems}, vol.~23, no.~7, pp. 8834--8845, 2021.

\bibitem{li2022cooperative}
K.~Li, J.~Wang, and Y.~Zheng, ``Cooperative formation of autonomous vehicles in
  mixed traffic flow: Beyond platooning,'' \emph{IEEE Transactions on
  Intelligent Transportation Systems}, vol.~23, no.~9, pp. 15\,951--15\,966,
  2022.

\bibitem{baldi2020establishing}
S.~Baldi, D.~Liu, V.~Jain, and W.~Yu, ``Establishing platoons of bidirectional
  cooperative vehicles with engine limits and uncertain dynamics,'' \emph{IEEE
  Transactions on Intelligent Transportation Systems}, vol.~22, no.~5, pp.
  2679--2691, 2020.

\bibitem{yu2023stability}
W.~Yu, D.~Ngoduy, X.~Hua, and W.~Wang, ``On the stability of a heterogeneous
  platoon-based traffic system with multiple anticipations in the presence of
  connected and automated vehicles,'' \emph{Transportation Research Part C:
  Emerging Technologies}, vol. 157, p. 104389, 2023.

\bibitem{zhao2020vehicle}
C.~Zhao, X.~Duan, L.~Cai, and P.~Cheng, ``Vehicle platooning with non-ideal
  communication networks,'' \emph{IEEE transactions on vehicular technology},
  vol.~70, no.~1, pp. 18--32, 2020.

\bibitem{halder2022stability}
K.~Halder, L.~Gillam, S.~Dixit, A.~Mouzakitis, and S.~Fallah, ``Stability
  analysis with lmi based distributed hinfinity controller for vehicle
  platooning under random multiple packet drops,'' \emph{IEEE Transactions on
  Intelligent Transportation Systems}, vol.~23, no.~12, pp. 23\,517--23\,532,
  2022.

\bibitem{dutta2020design}
R.~G. Dutta, Y.~Hu, F.~Yu, T.~Zhang, and Y.~Jin, ``Design and analysis of
  secure distributed estimator for vehicular platooning in adversarial
  environment,'' \emph{IEEE Transactions on Intelligent Transportation
  Systems}, vol.~23, no.~4, pp. 3418--3429, 2020.

\bibitem{yang2019multi}
T.~Yang, ``Multi-observer approach for estimation and control under adversarial
  attacks,'' Ph.D. dissertation, Doctoral thesis, Department of Electrical and
  Electronic Engineering, The University of Melbourne, 2019.

\bibitem{sontag2013mathematical}
E.~D. Sontag, \emph{Mathematical control theory: deterministic finite
  dimensional systems}.\hskip 1em plus 0.5em minus 0.4em\relax Springer Science
  \& Business Media, 2013, vol.~6.

\bibitem{willems2004deterministic}
J.~C. Willems, ``Deterministic least squares filtering,'' \emph{Journal of
  econometrics}, vol. 118, no. 1-2, pp. 341--373, 2004.

\bibitem{na2017adaptive}
J.~Na, G.~Herrmann, and K.~G. Vamvoudakis, ``Adaptive optimal observer design
  via approximate dynamic programming,'' in \emph{2017 American Control
  Conference (ACC)}.\hskip 1em plus 0.5em minus 0.4em\relax IEEE, 2017, pp.
  3288--3293.

\bibitem{chong2015parameter}
M.~S. Chong, D.~Ne{\v{s}}i{\'c}, R.~Postoyan, and L.~Kuhlmann, ``Parameter and
  state estimation of nonlinear systems using a multi-observer under the
  supervisory framework,'' \emph{IEEE Transactions on Automatic Control},
  vol.~60, no.~9, pp. 2336--2349, 2015.

\bibitem{armaghan2011design}
S.~Armaghan, A.~Moridi, and A.~K. Sedigh, ``Design of a switching pid
  controller for a magnetically actuated mass spring damper,'' in
  \emph{Proceedings of the World Congress on Engineering}, vol.~3, 2011, pp.
  6--8.

\bibitem{ref:Williams2007}
R.~L. Williams and D.~A. Lawrence, \emph{{Linear state-space control
  systems}}.\hskip 1em plus 0.5em minus 0.4em\relax Hoboken, NJ, USA: John
  Wiley {\&} Sons, 2007.

\bibitem{li2015design}
L.~Li, F.~Liao \emph{et~al.}, ``Design of a preview controller for
  discrete-time systems based on lmi,'' \emph{Mathematical Problems in
  Engineering}, vol. 2015, 2015.

\bibitem{yang2021secure}
T.~Yang and C.~Lv, ``A secure sensor fusion framework for connected and
  automated vehicles under sensor attacks,'' \emph{IEEE Internet of Things
  Journal}, vol.~9, no.~22, pp. 22\,357--22\,365, 2021.

\bibitem{naus2010string}
G.~J. Naus, R.~P. Vugts, J.~Ploeg, M.~J. van De~Molengraft, and M.~Steinbuch,
  ``String-stable cacc design and experimental validation: A frequency-domain
  approach,'' \emph{IEEE Transactions on vehicular technology}, vol.~59, no.~9,
  pp. 4268--4279, 2010.

\bibitem{klinge2009string}
S.~Klinge and R.~H. Middleton, ``String stability analysis of homogeneous
  linear unidirectionally connected systems with nonzero initial conditions,''
  2009.

\bibitem{an1997mechanical}
N.~An and S.~SI, ``Mechanical vibration and shock-evaluation of human exposure
  to whole-body vibration-part 1: General requirements,'' 1997.

\bibitem{papaioannou2023impact}
G.~Papaioannou, R.~Desai, and R.~Happee, ``The impact of body and head dynamics
  on motion comfort assessment,'' \emph{arXiv preprint arXiv:2307.03608}, 2023.

\bibitem{ko2021approach}
B.~Ko and S.~H. Son, ``An approach to detecting malicious information attacks
  for platoon safety,'' \emph{IEEE Access}, vol.~9, pp. 101\,289--101\,299,
  2021.

\bibitem{karmakar2021assessing}
G.~Karmakar, A.~Chowdhury, R.~Das, J.~Kamruzzaman, and S.~Islam, ``Assessing
  trust level of a driverless car using deep learning,'' \emph{IEEE
  Transactions on Intelligent Transportation Systems}, vol.~22, no.~7, pp.
  4457--4466, 2021.

\bibitem{nesterov1994interior}
Y.~Nesterov and A.~Nemirovskii, \emph{Interior-point polynomial algorithms in
  convex programming}.\hskip 1em plus 0.5em minus 0.4em\relax SIAM, 1994.

\bibitem{khalil2002nonlinear}
\BIBentryALTinterwordspacing
H.~Khalil, \emph{Nonlinear Systems}, ser. Pearson Education.\hskip 1em plus
  0.5em minus 0.4em\relax Prentice Hall, 2002. [Online]. Available:
  \url{https://books.google.nl/books?id=t_d1QgAACAAJ}
\BIBentrySTDinterwordspacing

\bibitem{ISO2631}
ISO2631, ``Mechanical vibration and shock—evaluation of human exposure to
  whole body vibration. part 1: General requirements,'' \emph{International
  Standard ISO 2631--1}, 1997.

\bibitem{taylor2022safety}
S.~J. Taylor, F.~Ahmad, H.~N. Nguyen, S.~A. Shaikh, and D.~Evans, ``Safety,
  stability and environmental impact of fdi attacks on vehicular platoons,'' in
  \emph{NOMS 2022-2022 IEEE/IFIP network operations and management
  symposium}.\hskip 1em plus 0.5em minus 0.4em\relax IEEE, 2022, pp. 1--6.

\end{thebibliography}
